\documentstyle[12pt,epsf,epsfig]{article}
\def\lsim{\mathrel{\rlap{\lower3pt\hbox{\hskip0pt$\sim$}}
    \raise1pt\hbox{$<$}}}         
\def\gsim{\mathrel{\rlap{\lower4pt\hbox{\hskip1pt$\sim$}}
    \raise1pt\hbox{$>$}}}         
\def\simlt{\mathrel{\raise.3ex\hbox{$<$\kern-.75em\lower1ex\hbox{$\sim$}}}}
\def\simgt{\mathrel{\raise.3ex\hbox{$>$\kern-.75em\lower1ex\hbox{$\sim$}}}}
\begin{document}

\begin{titlepage}


\begin{center}
\baselineskip25pt

\vspace{1cm}

{\Large\bf Renormalization Group Invariants in the MSSM and Its Extensions}
 \vspace{1cm}

Durmu{\c s} A. Demir
\vspace{0.3cm}

{\it Department of Physics, Izmir Institute of
Technology, IZTECH, Turkey, TR35430}

\end{center}
\vspace{1cm}
\begin{abstract}
We derive one-loop renormalization group (RG) invariant observables and analyze their phenomenological
implications in the MSSM and its $\mu$ problem solving extensions, U(1)$^{\prime}$ model and NMSSM. We show that
there exist several RG invariants in the gauge, Yukawa and soft-breaking sectors of each model. In general, RG
invariants are highly useful for projecting experimental data to messenger scale, for revealing correlations
among the model parameters, and for probing the mechanism that breaks supersymmetry. The Yukawa couplings and
trilinear soft terms in  U(1)$^{\prime}$ model and NMSSM do not form RG invariants though there exist approximate
invariants in low $\tan\beta$ domain. In the NMSSM, there are no invariants that contain the Higgs mass-squareds.
We provide a comparative analysis of RG invariants in  all three models and analyze their model-building and
phenomenological implications by a number of case studies.

\end{abstract}

\end{titlepage}

\section{Introduction}
The supersymmetric models provide an elegant solution to the gauge
hierarchy problem in a genuinely perturbative way for all scales right up to the Planckian territory. The
perturbative nature of the model allows one to relate measurements at the electroweak scale to physics at ultra
high energies. This communication between the infrared (IR) and ultraviolet (UV) regimes proceeds with the
renormalization group (RG) flow of the lagrangian parameters. Indeed, various phenomena central to supersymmetry
phenomenology $e.g.$ gauge coupling unification \cite{unif}, radiative electroweak breaking \cite{radi},
induction of flavor structures \cite{flavor} even for flavor-blind soft terms are pure renormalization effects.

Projection of experimental data to ultra high energies requires solving renormalization group equations (RGEs)
for parameters of the model. This projection, however, is generally complicated by the coupled nature of RGEs in
that measurement of a set of parameters at low scale cannot directly be rescaled to ultra high energies due to
leakage of other, possibly unknown, quantities. Therefore, in course of fitting a given model to laboratory and
astrophysical data it would be advantageous to have as much information as possible about correlations among the
parameters. Concerning this, a highly useful tool is provided by the RG-invariant observables. Indeed, such
quantities prove highly useful not only for projecting the experimental data to high energies but also for
deriving certain sum rules which enable fast consistency checks of the model \cite{kobayashi1, kobayashi2,
jones}. However, it should be kept in mind that, even the RG-invariant observables cannot be guaranteed to work
perfectly because ($i$) the RG invariance holds at a given loop order and it is generically disrupted by higher
loop effects (In general, scale and conformal invariances imply each other \cite{scale}, and superconformal group
involves both scale invariance and a continuous $R$ symmetry with correlated charges \cite{superconf}. Therefore,
in softly broken supersymmetric theories, where $R$ invariance is explicitly broken, the RG invariance, if any,
cannot be an all-order effect.), ($ii$) the RGEs get modified at sparticle thresholds so that what equations must
be used is not known {\it a priori}, and ($iii$) parameters with nontrivial flavor structures typically do not
exhibit RG-invariant combinations (this can be tied up to the fact that scale invariance puts strong constraints
on flavor structures of rigid and soft parameters of the theory \cite{ann}). The flavor mixings and sparticle
thresholds disrupt RG equations and associated invariants already at one loop order. On the other hand,
modification in a given RG invariant due to higher loop contributions is of order one less loop factor. For
instance, disruption of a one-loop RG invariant by two loop effects is of one loop order. Below we will restrict
our analysis to one loop RGEs with no flavor mixings in fermion sector.

This work is devoted to derivations and analyses of RG-invariant observables in the minimal supersymmetric model
(MSSM) and its $\mu$-problem solving minimal extensions $i.e.$ the next-to-minimal supersymmetry (NMSSM) and its
gauged version U(1)$^{\prime}$ model. Indeed, MSSM suffers from the naturalness problem associated with the Dirac
mass of Higgsinos. This mass parameter, $\mu$, is nested in the superpotential of the theory, and hence, its
scale is not controlled by the mechanism that breaks supersymmetry \cite{muprob}. Consequently, it is necessary
to find a mechanism for stabilizing $\mu$ to the electroweak scale. In fact, U(1)$^{\prime}$ model and NMSSM both
provide a dynamical solution to the problem by inducing $\mu$ via the VEV of an MSSM-singlet chiral superfield.
The U(1)$^{\prime}$ models are extensions of the MSSM by both an MSSM singlet and an additional Abelian symmetry
U(1)$^{\prime}$ \cite{u1prime,cvetic}. On the other hand, NMSSM has the same gauge group as MSSM yet its spectrum
contains a pure gauge singlet \cite{nmssm}. One here notes that a ${\rm Tev}$ scale U(1)$^{\prime}$ symmetry or
NMSSM are not necessarily the only solutions to the $\mu$ problem. Indeed, modification of the Kahler potential
by operators of the form $M_{Pl}^{-1}\, \widehat{z}^{\dagger} \widehat{H}_u \widehat{H}_d$, $\widehat{z}$ being a
hidden sector field,  generates the $\mu$ parameter at the right scale provided that theory possesses a global
Peccei-Quinn or continuous R invariance to forbid a bare $\mu$ parameter to appear in the superpotential
\cite{gm} (see also the related scenarios in \cite{casas} and \cite{ma}). Apart from this, the mechanism proposed
in \cite{kim} provides a simultaneous solution to the $\mu$ problem and the scale of supersymmetry breaking
within the supergravity framework by constructing explicit models of the hidden sector.

The RG-invariant observables and their phenomenological implications have already been analyzed in various
contexts. In addition to discussions in \cite{kobayashi1,kobayashi2,jones} there have been studies of the RG
invariants \cite{rginv} and resulting sum rules \cite{ramondmartin} within supersymmetric gauge theories and
certain string-inspired soft terms. In this work, we will provide a comparative analysis of the RG invariants in
the MSSM and its $\mu$ problem solving extensions. The RGEs for U(1)$^{\prime}$ model had been first given in
\cite{cvetic}. Here we generalize them to finite bottom and tau Yukawas. They are listed in Appendix A. The RGEs
for NMSSM had been given in \cite{king}, and we rederive and list them in Appendix B, for completeness. In
appendices we also discuss limiting cases where U(1)$^{\prime}$ and NMSSM RGEs reduce to those of the MSSM
\cite{rge}.

We start our analysis, in Sec. 2 below, by first describing the common part of
all three models $i.e.$ the sfermion sector. Then we derive RG-invariant observables and
discuss their phenomenological implications for the MSSM in Sec.3, for U(1)$^{\prime}$ model
in Sec. 4, and finally for the NMSSM in Sec.5. In Sec.6 we conclude the work.

\section{Generalities}
For the three supersymmertic models we will discuss the fermion sector is common whereas the Higgs and gauge
sectors vary from model to model. In general, one can write
\begin{eqnarray}
\widehat{W} = \widehat{W}_{Higgs} + \widehat{W}_{fermion}
\end{eqnarray}
where the superpotential of the fermion sector is given by
\begin{eqnarray}
\label{rigid-fermion} \widehat{W}_{fermion} = \widehat{U} {\bf Y_u} \widehat{Q} \widehat{H}_u + \widehat{D} {\bf
Y_d} \widehat{Q} \widehat{H}_d + \widehat{E} {\bf Y_e} \widehat{L} \widehat{H}_d
\end{eqnarray}
which encodes the Yukawa couplings ${\bf Y_{u,d,e}}$ (of up quarks, down quarks and of leptons) each being a
$3\times 3$ non-hermitian matrix in the space of fermion flavors. The fermion masses are induced by the vacuum
expectation values of the Higgs doublets $\widehat{H}_u$ and $\widehat{H}_d$, common to all three models. In
$\widehat{W}_{fermion}$ the left-handed quarks are arranged in SU(2)$_L$ doublets $\widehat{Q}$ and the
left-handed leptons in SU(2)$_L$ doublet $\widehat{L}$. On the other hand,  the left-handed anti up and anti down
quarks are represented by $\widehat{U}$ and $\widehat{D}$, respectively. Finally, $\widehat{E}$ collects
left-handed anti leptons\footnote{The neutrino masses and hence the requisite superfields (heavy Majorana
neutrinos or light right-handed neutrinos) are not incorporated in supersymmetric models under discussion.}.

The breakdown of supersymmetry is parameterized by various soft terms belonging to Higgs, gaugino and scalar
fermion sectors (see, $e.g.$ the review volume \cite{kane}):
\begin{eqnarray}
{\cal{L}}_{soft} = {\cal{L}}_{soft}^{Higgs}+ {\cal{L}}_{soft}^{gaugino}+{\cal{L}}_{soft}^{sfermion}
\end{eqnarray}
whose sfermion part reads as
\begin{eqnarray}
 \label{soft-fermion}
-{\cal{L}}_{soft}^{sfermion}&=&\widetilde{Q}^{\dagger} {\bf m_Q^2} \widetilde{Q} + \widetilde{U} {\bf m_U^2}
\widetilde{U}^{\dagger} + \widetilde{D} {\bf m_D^2} \widetilde{D}^{\dagger} + \widetilde{L}^{\dagger} {\bf m_L^2}
\widetilde{L} + \widetilde{E} {\bf m_E^2} \widetilde{E}^{\dagger}\nonumber\\
&+&\left[\widetilde{U} {\bf Y_u^A} \widetilde{Q} {H}_u + \widetilde{D} {\bf Y_d^A}\widetilde{Q} {H}_d +
\widetilde{E} {\bf Y_e^A} \widetilde{L} {H}_d + \mbox{h.c.}\right]
\end{eqnarray}
where ${\bf Y_{u,d,e}^A}$, like Yukawas themselves,  are non-hermitian flavor matrices whereas the sfermion
mass-squareds ${\bf m_{Q,\dots,E}^2}$ are all hermitian.

The interactions contained in (\ref{rigid-fermion}) and (\ref{soft-fermion}) exhibit mixings of various
flavors in both rigid and soft terms. As mentioned in Introduction, we focus only on the flavor-diagonal
interactions due to the fact that flavor mixings generically prohibit the construction of RG invariants except
for those parameters which depend on traces or determinants of the flavor matrices. Moreover, one recalls that
there is a certain degree of correlation between conformal invariance and flavor violation in that the former can
put stringent constraints on the latter \cite{ann}. Consequently, we switch off flavor mixings in all rigid and
soft parameters to obtain
\begin{eqnarray}
\label{approx} &&{\bf Y_{u,d,e}} \rightarrow \mbox{diag.}\left(0, 0, h_{t, b, \tau}\right),\;\,\nonumber\\ &&{\bf
Y_{u,d,e}^A} \rightarrow
\mbox{diag.}\left(0, 0, h_{t, b, \tau} A_{t, b, \tau}\right),\;\,\nonumber\\
&& {\bf m_Q^2} \rightarrow \mbox{diag.}\left(m_{\widetilde{u}_L}^2, m_{\widetilde{c}_L}^2, m_{\widetilde{t}_L}^2\right),\;\,\nonumber\\
&&{\bf m_L^2}
\rightarrow
\mbox{diag.}\left(m_{\widetilde{e}_L}^2, m_{\widetilde{\mu}_L}^2, m_{\widetilde{\tau}_L}^2\right),\;\,\nonumber\\
&& {\bf m_U^2} \rightarrow \mbox{diag.}\left(m_{\widetilde{u}_R}^2, m_{\widetilde{c}_R}^2, m_{\widetilde{t}_R}^2\right),\;\,\nonumber\\
&&{\bf m_D^2}
\rightarrow \mbox{diag.}\left(m_{\widetilde{d}_R}^2, m_{\widetilde{s}_R}^2, m_{\widetilde{b}_R}^2\right),\;\,\nonumber\\ &&{\bf m_E^2}
\rightarrow
\mbox{diag.}\left(m_{\widetilde{e}_R}^2, m_{\widetilde{\mu}_R}^2, m_{\widetilde{\tau}_R}^2\right),
\end{eqnarray}
where $m_{\widetilde{u}_L}^2=m_{\widetilde{d}_L}^2$,  $m_{\widetilde{c}_L}^2=m_{\widetilde{s}_L}^2$ and
$m_{\widetilde{t}_L}^2=m_{\widetilde{b}_L}^2$ by gauge invariance. Note that
light fermion Yukawa couplings are totally neglected. This reduction scheme for flavor mixings sets up the
notation and framework for the fermion sector. The gauge and Higgs sectors differ from model to model, and they
will be discussed in detail in the following sections.

Another model-independent aspect to be noted concerns IR and UV boundaries of the RGEs. For all three
supersymmetric models of interest, we neglect  modifications in the particle spectrum and RGEs coming from
decoupling of the heavy fields. In other words, we assume that all soft masses are approximately equal to
$M_{SUSY} \sim 1\, {\rm TeV}$ in logarithmic sense. This scale sets up the IR boundary for exact supersymmetric
RG flow. The UV boundary lies just beneath the scale of string territory, and we will take it to be the scale of
gauge coupling unification in the MSSM: $M_{GUT} \sim 10^{16}\, {\rm GeV}$. Therefore, in our framework, the RG
invariance of a given quantity means its scale independence in between the IR and UV scales above. In what
follows, we judiciously combine the RGEs of individual quantities until we arrive at a RG-invariant observable
within one loop accuracy. In general, there is no guarantee of maintaining RG invariance of a given quantity at
higher loop levels.

\section{The RG Invariants in the MSSM}
The MSSM is based on SU(3)$_c\times$SU(2)$_L\times$U(1)$_Y$ gauge group with respective gauge couplings $g_3$,
$g_2$ and $g_1$. The Higgs sector is spanned by $\widehat{H}_u$ and $\widehat{H}_d$ so that
\begin{eqnarray}
\label{rigid-Higgs-MSSM} \widehat{W}_{Higgs} = \mu \widehat{H}_u \widehat{H}_d
\end{eqnarray}
and
\begin{eqnarray}
 \label{soft-Higgs}
-{\cal{L}}_{soft}^{Higgs}&=& m_{H_u}^2 H_u^{\dagger} H_u + m_{H_d}^2 H_d^{\dagger} H_d + \left[\mu B {H}_u {H}_d
+ \mbox{h.c.}\right],\nonumber\\
-{\cal{L}}_{soft}^{gaugino}&=& \frac{1}{2}\sum_{a=3,2,1} \left[ M_{a} \lambda_{a} \lambda_{a} +
\mbox{h.c.}\right]
\end{eqnarray}
where $M_a$ is the gaugino mass.

By using one-loop RGEs (the RGEs in the MSSM have been computed up to two (partially up to three) loops order in
\cite{rge}) within the MSSM one can derive a number of invariants. Several RG invariants are listed in Table
\ref{table1}. The invariant $I_1$ correlates the gauge couplings with arbitrary constants $c_1$ and $c_2$. The
constants, however, can be related by using the values of gauge couplings at $M_{SUSY}$ and $M_{GUT}$:
\begin{eqnarray}
\label{gcoupleqn} \frac{c_1}{c_2} = -\frac{5}{3} \frac{g_1(M_{SUSY})^2}{g_2(M_{SUSY})^2}\left( \frac{g_0^2
g_2(M_{SUSY})^2 + 3 g_0^2 g_3(M_{SUSY})^2 - 4 g_2(M_{SUSY})^2 g_3(M_{SUSY})^2} {11 g_0^2 g_1(M_{SUSY})^2 + 5
g_0^2 g_3(M_{SUSY})^2 - 16 g_1(M_{SUSY})^2 g_3(M_{SUSY})^2}\right)
\end{eqnarray}
where $g_0$ is the common value of the gauge couplings at the unification scale $M_{GUT}$.

The second invariant $I_2$ in Table \ref{table1} correlates $\mu$ parameter with gauge and Yukawa couplings. From
this one can determine $\mu$ at any scale $Q\in \left[M_{SUSY}, M_{GUT}\right]$:
\begin{eqnarray}
\label{mueqn} \mu(Q_2) &=& \mu(Q_1) \left(\frac{h_t(Q_2)}{h_t(Q_1)}\right)^{\frac{27}{61}}
\left(\frac{h_b(Q_2)}{h_b(Q_1)}\right)^{\frac{21}{61}}
\left(\frac{h_{\tau}(Q_2)}{h_{\tau}(Q_1)}\right)^{\frac{10}{61}}\nonumber\\ &\times&
\left(\frac{g_3(Q_1)}{g_3(Q_2)}\right)^{\frac{256}{183}} \left(\frac{g_2(Q_1)}{g_2(Q_2)}\right)^{\frac{9}{61}}
\left(\frac{g_1(Q_2)}{g_1(Q_1)}\right)^{\frac{73}{2013}}
\end{eqnarray}
which makes it manifest that $\mu$ at any scale $Q$ depends on the strong coupling $g_3$ although its RGE does
not exhibit such a direct dependence at all. This exemplifies one interesting aspect of the RG invariants: they
make various otherwise implicit dependencies explicit. By putting $Q_2=M_{SUSY}$ and $Q_1=M_{GUT}$ one finds that
the ratio $\mu(M_{SUSY})/\mu(M_{GUT})$, which is one of the most crucial factors (together with the gluino mass)
that determine the amount of fine-tuning needed to achieve the correct value of the $Z$ boson mass, is entirely
determined by the interplay between the IR and UV values of the rigid parameters. In particular, (\ref{mueqn})
suggests that $\mu(M_{SUSY})/\mu(M_{GUT})$ decreases with increasing $\tan\beta$: $\mu(M_{SUSY})/\mu(M_{GUT})
\simeq 0.96$ for $\tan\beta=5$ and $\simeq 0.3$ for $\tan\beta=60$. Indeed, this ratio is governed mainly by
$g_3$ at low $\tan\beta$ and by $g_3$, $h_b$ and $h_{\tau}$ for $\tan\beta\sim m_t/m_b$. Therefore, the
sensitivity of $M_Z$ to $\mu(M_{GUT})$ is greatly reduced at large $\tan\beta$ which itself requires a great deal
of fine-tuning to achieve though \cite{randall} (see \cite{tata} for a discussion of the fine-tuning problems in
large $\tan\beta$ domain when radiative corrections to Higgs potential are taken into account).

\begin{table}[h]
\begin{center}
\begin{tabular}{|c||c|}
  \hline
  Number  & RG Invariant \\ \hline \hline
  $I_1$  &  $\frac{c_1}{g_1^2}+\frac{c_2}{g_2^2}+\frac{33 c_1 +5 c_2}{15 g_3^2}$ \\  \hline
  $I_2$  &  $\mu\, \left(\frac{g_2^9\, g_3^{256/3}}{h_t^{27}\, h_b^{21}\, h_{\tau}^{10}\,
g_1^{73/33}}\right)^{1/61}$\\ \hline\hline $I_3$  & $\frac{M_{a}}{g_{a}^2}\; (a=1,2,3) $ \\
\hline $I_4$  & $ B - \frac{27}{61} A_t
- \frac{21}{61} A_b - \frac{10}{61} A_{\tau} - \frac{256}{183} M_3 - \frac{9}{61} M_2 + \frac{73}{2013} M_1$\\
\hline\hline $I_5$  & $m_{\widetilde{\tau}_R}^2 - 2 m_{\widetilde{\tau}_L}^2 - 3 \left|M_2\right|^{2} + \frac{1}{11} \left|M_1\right|^2-
\frac{6}{13} S$\\ \hline $I_6$    & $ m_{H_u}^2 - \frac{3}{2} m_{\widetilde{t}_R}^2 + \frac{4}{3} \left|M_3\right|^{2} +
\frac{3}{2} \left|M_2\right|^{2} - \frac{5}{66} \left|M_1\right|^2 - \frac{9}{26}S$\\\hline $I_7$   & $m_{H_d}^2
- \frac{3}{2} m_{\widetilde{b}_R}^2 - m_{\widetilde{\tau}_L}^2 + \frac{4}{3} \left|M_3\right|^{2} - \frac{1}{33} \left|M_1\right|^2 +
\frac{3}{26} S$\\ \hline $I_8$  & $m_{\widetilde{t}_R}^2 + m_{\widetilde{b}_R}^2 - 2 m_{\widetilde{t}_L}^2 - 3 \left|M_2\right|^2 +
\frac{1}{11}
\left|M_1\right|^2 + \frac{2}{13} S$\\ \hline $I_9$  & $m_{\widetilde{u}_L}^2+\frac{1}{198} \left|M_1\right|^2 + \frac{3}{2}
\left|M_2\right|^2 - \frac{8}{9} \left|M_3\right|^2 - \frac{1}{26} S$\\ \hline $I_{10}$  &
$m_{\widetilde{u}_R}^2+\frac{8}{99} \left|M_1\right|^2 - \frac{8}{9} \left|M_3\right|^2 + \frac{2}{13} S$\\ \hline
$I_{11}$  & $m_{\widetilde{d}_R}^2+\frac{2}{99} \left|M_1\right|^2 - \frac{8}{9} \left|M_3\right|^2 - \frac{1}{13} S$\\
\hline $I_{12}$  & $m_{\widetilde{e}_L}^2+\frac{1}{22} \left|M_1\right|^2 - \frac{3}{2} \left|M_2\right|^2 + \frac{3}{26}
S$\\ \hline $I_{13}$  & $m_{\widetilde{e}_R}^2+\frac{2}{11} \left|M_1\right|^2- \frac{3}{13} S$\\ \hline
 $I_{14}$  & $m_{H_u}^2 + m_{H_d}^2 - 3 m_{\widetilde{t}_L}^2 - m_{\widetilde{\tau}_L}^2
+ \frac{8}{3} \left|M_3\right|^2 -3 \left|M_2\right|^2 + \frac{2}{11} \left|M_1\right|^2$\\ \hline $I_{15}$  &
$m_{H_d}^2 - \frac{3}{2} m_{\widetilde{b}_R}^2 - \frac{3}{2} m_{\widetilde{\tau}_L}^2 + \frac{1}{4} m_{\widetilde{\tau}_R}^2 + \frac{4}{3}
\left|M_3\right|^{2} - \frac{3}{4} \left|M_2\right|^{2} - \frac{1}{132} \left|M_1\right|^2$\\ \hline $I_{16}$  &
$2 m_{\widetilde{u}_L}^2+m_{\widetilde{u}_R}^2+m_{\widetilde{d}_R}^2 + \frac{1}{9} \left|M_1\right|^2 + 3 \left|M_2\right|^2 - \frac{32}{9}
\left|M_3\right|^2$\\ \hline $I_{17}$  & $m_{\widetilde{e}_L}^2+ \frac{1}{2} m_{\widetilde{e}_R}^2 + \frac{3}{22} \left|M_1\right|^2 -
\frac{3}{2} \left|M_2\right|^2$\\ \hline
\end{tabular}
\end{center}
\caption{The RG invariant combinations of rigid and soft parameters in the MSSM ($c_1$ and $c_2$ in $I_1$ are
arbitrary constants). Note that invariants pertaining to the first and second generations generically involve a
single sfermion mass-squared since trilinear couplings do not contribute to their RGEs. In a sense, these are
'fundamental' invariants derived directly from the RG flows of relevant parameters. For obtaining RG invariants
containing a specific set of parameters it is necessary to form appropriate combinations of these tabulated ones,
as exemplified in the text by a couple of case studies. The quantity $S$ appearing in some of the invariants is
defined in equation (\ref{seqn}).} \label{table1}
\end{table}

The third line of Table \ref{table1} shows that the ratio of the gaugino mass to fine structure constant of the
same group is an RG invariant. This invariance property guarantees that
\begin{eqnarray}
\label{maeqn} M_a(Q_2) = M_a(Q_1) \left(\frac{g_a(Q_2)}{g_a(Q_1)}\right)^2
\end{eqnarray}
so that knowing two of the gaugino masses at a scale $Q$ suffices to know the third if gauge coupling unification
holds -- an important aspect to check directly the minimality of the gauge structure using the experimental data.
This very relation also shows that $M_3(M_{SUSY})/M_3(M_{GUT})$ is much larger
$M_{1,2}(M_{SUSY})/M_{1,2}(M_{GUT})$ due to asymptotic freedom. In fact, in minimal superhravity for instance,
typically gluino is the first superpartner to decouple from the light spectrum.

The fourth line of Table \ref{table1} correlates Higgs bilinear soft term $B$ with trilinear couplings and
gaugino masses. Among various possibilities, by using this invariant one can express, for instance, $B$ at any
scale $Q$ in terms of other dimension-one soft masses:
\begin{eqnarray} \label{beqn}
B(Q_2) &=& B(Q_1)+\frac{27}{61} \left(A_t(Q_2)-A_t(Q_1)\right) + \frac{21}{61}
\left(A_b(Q_2)-A_b(Q_1)\right)\nonumber\\ &+&\frac{10}{61} \left(A_{\tau}(Q_2)-A_{\tau}(Q_1)\right)
+\frac{256}{183} M_3(Q_1) \left( \frac{g_3(Q_2)^2}{g_3(Q_1)^2} - 1\right)\nonumber\\ &+& \frac{9}{61} M_2(Q_1)
\left( \frac{g_2(Q_2)^2}{g_2(Q_1)^2} - 1\right) -\frac{73}{2013} M_1(Q_1) \left( \frac{g_1(Q_2)^2}{g_1(Q_1)^2} -
1\right)
\end{eqnarray}
after using (\ref{maeqn}). This equation expresses the IR value of the $B$ parameter in terms of the IR and UV
values of the gaugino masses and trilinear couplings. The RGE of the $B$ parameter does not depend on the gluino
mass explicitly (the dependence comes through the trilinear couplings); however, (\ref{beqn}) exhibits a rather
strong dependence on $M_3$: for $Q_2=M_{SUSY}$ and $Q_1=M_{GUT}$ the gluino contribution equals $2.6
M_3(M_{GUT})$ which is much larger than other contributions (except possibly the GUT scale value of $B$). This
very fact proves the power of forming RG invariant observables as they make indirect effects manifest.

Having completed the discussion of the rigid and dimension-one soft parameters of the theory, we now start
analyzing the scale-invariant combinations of the scalar mass-squareds. They are listed in Table \ref{table1}
starting from line 5. The RGEs of the soft mass-squareds depend on the quantity \cite{rge}
\begin{eqnarray}
\label{seqn} S =\mbox{Tr}\left[m^2 Y\right] &=&  m_{H_u}^2 - m_{H_d}^2 + \left(m_{\widetilde{t}_L}^2 - m_{\widetilde{t}_R}^2\right) +
\left(m_{\widetilde{c}_L}^2 - m_{\widetilde{c}_R}^2\right) + \left(m_{\widetilde{u}_L}^2 - m_{\widetilde{u}_R}^2\right)\nonumber\\ &+&
\left(m_{\widetilde{b}_R}^2 -
m_{\widetilde{t}_R}^2\right) + \left(m_{\widetilde{s}_R}^2-m_{\widetilde{c}_R}^2\right) +
\left(m_{\widetilde{d}_R}^2-m_{\widetilde{u}_R}^2\right)\nonumber\\& -&
\left(m_{\widetilde{\tau}_L}^2 - m_{\widetilde{\tau}_R}^2\right) - \left(m_{\widetilde{\mu}_L}^2 - m_{\widetilde{\mu}_R}^2\right) -
\left(m_{\widetilde{e}_L}^2 -
m_{\widetilde{e}_R}^2\right)
\end{eqnarray}
which comprises all of the soft mass-squareds. This quantity identically vanishes if they are strictly universal
at some given scale since then ${\mbox Tr}\left[m^2 Y\right] = m^2 {\mbox Tr}\left[Y\right] \equiv 0$ thanks to
the absence of the gravitational anomaly. As the explicit solution
\begin{eqnarray}
S(Q_2) = \left(\frac{g_1(Q_2)}{g_1(Q_1)}\right)^{26/33}\, S(Q_1)
\end{eqnarray}
also suggests, $S(Q)$ vanishes at all scales if it does so at some given scale. That the universality of the soft
mass--squareds,
\begin{eqnarray}
\label{univ}
m_{H_u}^2 = m_{H_d}^2= \cdots =  m_{\widetilde{\tau}_R}^2 =m_0^2,
\end{eqnarray}
renders theory $S$--free is important in that experimental tests of whether $S$ is vanishing or not can give
important information on if soft masses unify at ultra high energies. This universality scheme, when supplemented
by $A_t(M_{GUT})=A_b(M_{GUT})=A_{\tau}(M_{GUT})= A_0$ and $M_3(M_{GUT})=M_2(M_{GUT})=M_1(M_{GUT}) =M$, leads one
to the minimal supergravity configuration. One further notes that, the RGE of a scalar $\phi$ senses $S$ via the
contribution $Y_{\phi} (3/5) g_1^2 S$; however, the RG invariants of the soft mass-squareds depend on $S$ without
any $g_1^2$ dressing.

The RG-invariant combinations of the soft mass-squareds can therefore be analyzed in two groups: those that are
sensitive to $S$ (lines 5-13 of Table \ref{table1}) and those that are insensitive to $S$ (lines 14-17 of Table
\ref{table1}). Clearly, one can construct new invariants by combining these available ones. An accurate enough
measurement (presumably at LHC $\oplus$ ILC) of (all or part of the) soft mass-squareds will serve both as a
testing ground for the internal consistency of the model and a as tool for probing the ultra high energy behavior
(whether it is minimal supergravity or not within experimental error bounds) \cite{ramondmartin,spect}. More
specifically, by using these invariants one can ($i$) test the internal consistency of the model while fitting to
the experimental data; ($ii$) rehabilitate poorly known parameters supplementing the well-measured ones; ($iii$)
determine what kind of supersymmetry breaking mechanism is realized in Nature; and finally ($iv$) separately
examine the UV scale configurations of the trilinear couplings as they do not explicitly contribute to these
invariants. We now want to illustrate some of these useful aspects of these invariants by studying a few
interesting cases:
\begin{itemize}
\item The RG invariants provide useful constraints on the low-energy parameter space. For illustrating this point
let as consider, for example, the RG invariant $I_8$ in Table \ref{table1}. If the universality constraint
(\ref{univ}) holds at some scale $Q_1$ then one finds
\begin{eqnarray}
\label{split} m_{\widetilde{t}_R}^2(Q_2) + m_{\widetilde{b}_R}^2(Q_2) - 2 m_{\widetilde{t}_L}^2(Q_2) &=& 3 \left|M_2(Q_1)\right|^2
\left[\left(\frac{g_2(Q_2)}{g_2(Q_1)}\right)^4 -1\right]\nonumber\\ &-&  \frac{1}{11} \left|M_1(Q_1)\right|^2
\left[\left(\frac{g_1(Q_2)}{g_1(Q_1)}\right)^4 -1\right]
\end{eqnarray}
after using equation (\ref{maeqn}). This equality establishes a relation between the stop and sbottom masses
right at the scale of measurement in a way involving the UV values of the gaugino masses. Despite this, however,
it shows that $\widetilde{t}_L$--$\widetilde{t}_R$ plus $\widetilde{b}_L$--$\widetilde{b}_R$ mass splittings are
entirely controlled by the isospin and hypercharge gaugino masses rather than the gluino mass. This is an
important aspect as it significantly reduces sensitivity to the UV scale values of the gaugino masses. Indeed, by
taking $Q_2=M_{SUSY}$ and $Q_1=M_{GUT}$ the right-hand side of (\ref{split} reduces to $-0.97 M_2(M_{GUT})^2 +
0.08 M_1(M_{GUT})^2$ which does not exhibit any pronounced sensitivity to GUT scale gaugino masses (unlike, for
instance, (\ref{beqn})). One possible application of (\ref{split}) among many one can consider is that it
establishes a relation between the stop and sbottom mixing angles
\begin{eqnarray}
\label{angles} \sum_{f=t,b} \frac{ m_f |A_f - \mu^{\star} R_f|}{\tan 2 \theta_{\widetilde{f}}} &\simeq& -
 0.5 M_{2}(M_{GUT})^2 + 0.04 M_1(M_{GUT})^2 - 0.04 M_Z^2
\end{eqnarray}
where $R_f = \cot \beta (\tan\beta)$ for $f=t (b)$. In estimating the right hand side we took $\cos 2\beta \simeq
-1$ in accord with the LEP bounds which prefer fairly large values of $\tan\beta$. This simple formula may serve
as a constraint in simulating the supersymmetric parameter space as the experimental data accumulate. So far we
have assumed that the theory is $S$-free. What if it is not? In this case one automatically obtains a direct
sensitivity to all soft mass-squareds and neither (\ref{split}) nor (\ref{angles}) can provide a signal as clean
as in the universal case.

\item The RG invariants in Table \ref{table1} can be combined to obtain new invariants that involve solely the
scalar mass-squareds in the theory. For instance, by taking $M_{GUT}$ to be the UV scale with universal scalar
masses the Higgs soft masses can be expressed as
\begin{eqnarray}
\label{higgses} m_{H_u}^2(Q) &=& \frac{7}{12} m_0^2 + \frac{5}{12} m_{\widetilde{e}_R}^2(Q) + m_{\widetilde{u}_L}^2(Q) -
\frac{21}{12} m_{\widetilde{u}_R}^2(Q) - \frac{3}{4} m_{\widetilde{d}_R}^2(Q) + \frac{3}{2} m_{\widetilde{t}_R}^2(Q) \nonumber\\
m_{H_d}^2(Q) &=& - \frac{15}{4} m_0^2 -\frac{1}{4} m_{\widetilde{e}_R}^2(Q) -3 \left(m_{\widetilde{u}_L}^2(Q)-\frac{1}{2}
m_{\widetilde{u}_R}^2(Q)\right)\nonumber\\ &-& 3 \left(m_{\widetilde{t}_L}^2(Q) - \frac{1}{2} m_{\widetilde{t}_R}^2(Q) - \frac{3}{4}
m_{\widetilde{b}_R}^2(Q)\right) + \frac{3}{2} \left(m_{\widetilde{\tau}_L}^2(Q)-\frac{1}{6} m_{\widetilde{\tau}_R}^2(Q)\right)
\end{eqnarray}
which serve as a sum rule to correlate various soft masses with no contributions from those of the gauginos.
These expressions determine Higgs soft mass-squareds in terms of the sfermion masses and the universal scalar
mass at the GUT scale. Clearly, for determining the Higgs soft mass-squareds it is necessary to know soft masses
of all three generations of sfermions if the gaugino sector is to be skipped ( compare (\ref{higgses}) with the
invariants $I_6$, $I_7$ and $I_{14}$ in Table \ref{table1}).

The RG invariant $I_{13}$ can prove useful for determining $S$ from a minimal number of measurements. Indeed,
measuring $m_{\widetilde{e}_R}$ and $M_1$ at two distinct scales and taking the difference determines if $S$ is  vanishing
(if scalar masses attain a universal scheme as in (\ref{univ})) or not. However, given that a measurement of
$M_1$ requires exploration of the neutralino sector, a more promising invariant is
\begin{eqnarray}
m_{\widetilde{u}_R}^2-m_{\widetilde{d}_R}^2 - \frac{1}{3} m_{\widetilde{e}_R}^2 + \frac{4}{13} S
\end{eqnarray}
which involves only the first family sfermions which are simultaneous eigenstates of mass, gauge, flavor and
chirality (and thus, their experimental identification could be easier than those of gauginos and third
generation sfermions which undergo non-negligible mixings).

\item The RG invariants are highly useful probes of the mechanism that breaks the supersymmetry. We illustrate
their discriminative power by examining three well-known supersymmetry breaking schemes: ($i$) no-scale
supergravity models \cite{noscale}, ($ii$) dilaton-dominated supersymmetry breaking \cite{soft} and ($iii$)
flux-induced soft terms \cite{flux}. The soft-breaking sectors of these models commonly exhibit the minimal
supergravity (constrained MSSM) configuration: $m_{H_u}^2 =$ $m_{H_d}^2= \cdots =$  $m_{\widetilde{\tau}_R}^2 =m_0^2$,
$A_t(M_{GUT})=$ $A_b(M_{GUT})=$ $A_{\tau}(M_{GUT})= A_0$ and $M_3(M_{GUT})=$ $M_2(M_{GUT})=$ $M_1(M_{GUT}) =M$.
However, correlations among the parameters vary from model to model so does the pattern of the RG invariants. In
no-scale supergravity $A_0 =B(M_{GUT})=m_0=0$, in dilaton domination $m_0 = M/\sqrt{3}$, $A_0=- M$ and
$B(M_{GUT}) = 2 M/\sqrt{3}$, and in fluxed MSSM $m_0 = M$, $A_0=- 3 M$ and $B(M_{GUT}) = -2 M$. The values of the
soft-sector invariants are displayed in Tables \ref{table2} and \ref{table3}. One notices that, the only
model-independent invariant is $I_8$ as it solely probes if soft mass-squareds and/or gaugino masses are
universal or not. The other invariants differ from model to model. In case one invariant, say $I_4$, is
determined by experiment with sufficient accuracy and if it agrees with predictions of a specific model, say
dilaton-dominated supersymmetry breaking, then all one has to do is to check if rest of the invariants (to be
determined as more and more data accumulate) agree with the experiment. In this sense, the results displayed in
Tables \ref{table2} and \ref{table3} (which can be expanded to include all possible breaking schemes found in
strings, supergravity, anomaly mediation, gauge mediation, etc.) can be used as a look up table for
checking/predicting which mechanism of supersymmetry breaking is favored or realized in Nature. Clearly, RG
invariance is not a requisite property for an observable to probe supersymmetry breaking sector; however, if it
is RG-invariant it is not necessary to integrate the RGEs and it is possible to use results of different
experiments without RG scaling (irrespective of if they are obtained from a reanalysis of LEP data or from
Tevatron or from LHC).

\begin{table}[h]
\begin{center}
\begin{tabular}{|c|c|c|c|c|c|c|c|c|c|c|c|c|c|c|}
  \hline
Mechanism & $I_4/M$ & $I_5/M^2$ & $I_6/M^2$ & $I_7/M^2$ & $I_8/M^2$ & $I_9/M^2$\\ \hline\hline No-scale&
-3040/2013 & -32/11 & 91/33 & 43/33 & -32/11 & 61/99\\ \hline Dilaton  & $2/\sqrt{3}$-1126/2013 & -107/33 &
57/22 & 53/66 & -32/11 & 94/99\\ \hline Fluxed & -7462/2013 & -43/11 & 149/66 & -13/66 & -32/11 & 160/99\\
\hline
\end{tabular}
\end{center}
\caption{The patterns of the RG invariants ($I_4$--$I_9$) within no-scale \cite{noscale}, dilaton domination
\cite{soft} and fluxed MSSM \cite{flux} supersymmetry breaking schemes.}
\label{table2}
\end{table}

\begin{table}[h]
\begin{center}
\begin{tabular}{|c|c|c|c|c|c|c|c|c|c|c|c|c|c|c|}
  \hline
Mechanism & $I_{10}/M^2$ & $I_{11}/M^2$ &
  $I_{12}/M^2$ & $I_{13}/M^2$ & $I_{14}/M^2$ & $I_{15}/M^2$ & $I_{16}/M^2$ & $I_{17}/M^2$\\ \hline\hline
No-scale&  -80/99 & -86/99 & -16/11 & 2/11 & -10/33 & 19/33 & -4/9 & -15/11
\\\hline Dilaton  &  -47/99 & -53/99
& -37/33 & 17/33 &
 -32/33 & -1/132 & 8/9 & -19/22\\ \hline
Fluxed & 19/99 & 13/99 & -5/11 & 13/11 & -76/33 & -155/132 & 32/9 & 3/22
\\\hline
\end{tabular}
\end{center}
\caption{The same as in Table \ref{table2} but for $I_{10}$- $I_{17}$.} \label{table3}
\end{table}

\item The RG invariants do have interesting implications also for chargino and neutralino sectors of the model.
For instance, from the product of two chargino masses one immediately finds that the quantity
\begin{eqnarray}
\label{char} \left(M_{\chi_1^\pm} M_{\chi_2^\pm} + M_W^2 \sin 2 \beta\right) \left(\frac{g_3^{256/3}}{h_t^{27}\,
h_b^{21}\, h_{\tau}^{10}\, g_2^{113}\, g_1^{73/33}}\right)^{1/61}
\end{eqnarray}
is an RG invariant observable of the model. Besides, sum of the neutralino masses can be shown to be expressible
in terms of the gauge couplings and gluino mass only, as already derived in \cite{ramondmartin}. In the
neutralino sector, one finds that
\begin{eqnarray}
\label{carpim} &&\frac{1}{M_{\chi_1^\pm} M_{\chi_2^\pm}} \left\{ \prod_{i=1}^{4} M_{\chi_i^0} - \tan^2\theta_W
\sin 2\beta M_W^2
\left( M_{\chi_1^\pm} M_{\chi_2^\pm} + \sin 2\beta M_W^2\right)\right\}\nonumber\\
&\times& \left(\frac{g_2^9\, g_3^{256/3}}{h_t^{27}\, h_b^{21}\, h_{\tau}^{10}\,  g_1^{4099/33}}\right)^{1/61}
\end{eqnarray}
is an RG invariant. These invariants can be useful for determining the scale dependence of certain parameters
from a few measured ones, and also they manifestly depict the correlation between the neutralino/chragino and
gauge/Yukawa sectors. Finally, one notes that under the rescalings
\begin{eqnarray}
\label{rescale} \widetilde{H}_{u,d} \rightarrow \left(\frac{g_2^9\, g_3^{256/3}}{h_t^{27}\, h_b^{21}\,
h_{\tau}^{10}\, g_1^{73/33}}\right)^{-1/122}\, \widetilde{H}_{u,d},\; \widetilde{B} \rightarrow g_1
\widetilde{B},\; \widetilde{W} \rightarrow g_2 \widetilde{W}
\end{eqnarray}
the neutralino and chargino mass matrices become completely scale-invariant except for SU(2)$_L\otimes$U(1)$_Y$
breaking terms which mix Higgsinos and gauginos. This property could be useful in calculating and interpreting
certain observables.
\end{itemize}

We have derived a number of RG invariants in the MSSM, and performed certain case studies for highlighting their
phenomenological relevance. These invariants, as also emphasized in Introduction, could be quite useful for
determining the origin of supersymmetry breaking, for testing the internal consistency of the model, and for
obtaining certain sum rules that enable the prediction of certain unknown parameters from the known ones.

\section{The RG Invariants in U(1)$^\prime$ Model}
The U(1)$^\prime$ extension of the MSSM is based on SU(3)$_c\times$SU(2)$_L\times$U(1)$_Y\times$U(1)$^\prime$
gauge group with respective gauge couplings $g_3$, $g_2$, $g_1$ and $g_1^{\prime}$. The Higgs sector is spanned
by $\widehat{H}_u$, $\widehat{H}_d$ and $\widehat{S}$ so that
\begin{eqnarray}
\label{rigid-Higgs-U1prime} \widehat{W}_{Higgs} = h_s \widehat{S} \widehat{H}_u \widehat{H}_d
\end{eqnarray}
is the unique superpotential comprising the Higgs superfields since U(1)$^{\prime}$ invariance ($i$) forbids the
appearance of a bare $\mu$ parameter (as in the MSSM superpotential (\ref{rigid-Higgs-MSSM})), and ($ii$) doeshat
not allow for additional terms such as $\widehat{S}^3$ (as in the NMSSM superpotential to be discussed in the
next section) \cite{u1prime,cvetic}. The full superpotential is obtained by adding (\ref{rigid-Higgs-U1prime}) to
the fermionic part given in (\ref{rigid-fermion}).

The soft-breaking terms pertaining to Higgs and gaugino sectors are given by
\begin{eqnarray}
 \label{soft-Higgs-U1prime}
-{\cal{L}}_{soft}^{Higgs}&=& m_{H_u}^2 H_u^{\dagger} H_u + m_{H_d}^2 H_d^{\dagger} H_d + m_S^2 S^{\dagger} S +
\left[ h_s A_s S {H}_u\cdot {H}_d + \mbox{h.c.}\right],\nonumber\\
-{\cal{L}}_{soft}^{gaugino}&=& \frac{1}{2}\sum_{a=3,2,1,1'} \left[ M_{a} \lambda_{a} \lambda_{a} +
\mbox{h.c.}\right]\,.
\end{eqnarray}
A comparison of the superpotential and soft-breaking terms with their MSSM counterparts (\ref{rigid-Higgs-MSSM})
and (\ref{soft-Higgs}) shows clearly the way the MSSM limit is reached. Indeed, below the U(1)$^{\prime}$
breaking scale the effective theory resembles the MSSM (it just resembles because, for instance, the neutralino
sector of the MSSM is extended by U(1)$^{\prime}$ gaugino and singlino $\widetilde{S}$ states) with the
parameters
\begin{eqnarray}
\label{effectives} \mu_{eff} \equiv h_s \langle S \rangle\,,\; \, \mu_{eff} B_{eff} \equiv h_s A_s \langle S
\rangle
 \end{eqnarray}
which are both stabilized at the weak scale as desired if the singlet develops a VEV $\langle S \rangle$ at the
same scale \cite{u1prime,cvetic}. In essence, as far as the Higgs sector is concerned, the naturalness problem
associated with the $\mu$ parameter of the MSSM is avoided as it is now generated dynamically by U(1)$^{\prime}$
breakdown \cite{muprob}.

\begin{table}[h]
\begin{center}
\begin{tabular}{|c||c|}
  \hline
  Number  & RG Invariant \\ \hline \hline
  $I_1^{\prime}$  &  $ \frac{3 c_3 -c_2 - (33/5) c_1}{(6 + \rho \lambda_H)\, {g_1^{\prime}}^2}
  +\frac{c_1}{g_1^2}+\frac{c_2}{g_2^2}+\frac{c_3}{g_3^2}$ \\  \hline
  $I_2^{\prime}$  & $\approx h_s\,  h_t^{3/7}\, h_b^{-39/7}\, h_{\tau}^{8/7}\, g_3^{64/7}\, g_2^{-9}\, g_1^{5/77}\,
{g_1^{\prime}}^{-a^{\prime}}$\\ \hline\hline $I_3^{\prime}$  & $\frac{M_{a}}{g_{a}^2}\; (a=1^{\prime}, 1,2,3) $ \\
\hline $I_4^{\prime}$  & $\approx A_s + \frac{3}{7} A_t - \frac{39}{7} A_b + \frac{8}{7} A_{\tau} + \frac{64}{7}
M_3 - 9 M_2 + \frac{5}{77} M_1 - a^{\prime} M_1^{\prime}$ \\ \hline\hline $I_5^{\prime}$  & $ I_5+ \frac{\rho}{6
+ \rho \lambda_H} (Q_E^2 - 2 Q_L^2) \left|M_1^{\prime}\right|^2 $\\ \hline $I_6^{\prime}$    & $ I_6 -
\frac{1}{2} m_S^2 + \frac{\rho}{6 +  \rho \lambda_H} (Q_{H_u}^2 - \frac{1}{2} Q_S^2 - \frac{3}{2} Q_U^2)
\left|M_1^{\prime}\right|^2 $\\\hline $I_7^{\prime}$ & $ I_7 - \frac{1}{2} m_S^2 + \frac{\rho}{6 +  \rho
\lambda_H} (Q_{H_d}^2 - \frac{1}{2} Q_S^2 - \frac{3}{2} Q_D^2 - Q_L^2) \left|M_1^{\prime}\right|^2 $\\ \hline
$I_8^{\prime}$  & $ I_8 + \frac{\rho}{6 +  \rho \lambda_H} (Q_{U}^2 + Q_D^2 - 2 Q_Q^2)
\left|M_1^{\prime}\right|^2 $\\\hline $I_9^{\prime}$  & $I_9 + \frac{\rho}{6 +  \rho \lambda_H} Q_Q^2
\left|M_1^{\prime}\right|^2 $\\ \hline $I_{10}^{\prime}$  & $I_{10} + \frac{\rho}{6 +  \rho \lambda_H} Q_{U}^2
\left|M_1^{\prime}\right|^2 $\\ \hline $I_{11}^{\prime}$  & $ I_{11} + \frac{\rho}{6 +  \rho \lambda_H}
Q_D^2\left|M_1^{\prime}\right|^2 $\\ \hline $I_{12}^{\prime}$  & $ I_{12} + \frac{\rho}{6 +  \rho \lambda_H}
Q_{L}^2 \left|M_1^{\prime}\right|^2 $\\ \hline $I_{13}^{\prime}$  & $ I_{13} + \frac{\rho}{6 +  \rho \lambda_H}
Q_{E}^2 \left|M_1^{\prime}\right|^2 $\\ \hline $I_{14}^{\prime}$  & $ I_{14} - m_{S}^2 + \frac{\rho}{6 +  \rho
\lambda_H} (Q_{H_u}^2 + Q_{H_d}^2 - Q_S^2 - 3 Q_Q^2 - Q_L^2) \left|M_1^{\prime}\right|^2 $\\ \hline
$I_{15}^{\prime}$ & $I_{15}- \frac{1}{2} m_S^2 + \frac{\rho}{6 +  \rho \lambda_H} (Q_{H_d}^2 - \frac{1}{2}
Q_{S}^2 - \frac{3}{2} Q_D^2 - \frac{3}{2} Q_L^2 + \frac{1}{4} Q_E^2) \left|M_1^{\prime}\right|^2 $\\ \hline
$I_{16}^{\prime}$  & $I_{16} + \frac{\rho}{6 +  \rho \lambda_H} (2 Q_{Q}^2 + Q_{U}^2 + Q_D^2)
\left|M_1^{\prime}\right|^2 $\\ \hline $I_{17}^{\prime}$  & $I_{17} + \frac{\rho}{6 +  \rho \lambda_H} (Q_{L}^2 +
\frac{1}{2} Q_{E}^2) \left|M_1^{\prime}\right|^2 $\\\hline
\end{tabular}
\end{center}
\caption{The RG invariant combinations of rigid and soft parameters in U(1)$^{\prime}$ models ( $c_{1,\dots,3}$
are arbitrary constants). Here $I_2^{\prime}$ and $I_4^{\prime}$ are approximate invariants derived in the text.
The invariants constructed from scalar mass-squareds are written in terms of the MSSM invariants in Table
\ref{table1}. The modifications are twofold: First, the MSSM invariants involving Higgs mass-squareds are shifted
by the singlet mass-squared. Next, each invariant receives new contributions proportional to
$\left|M_1^{\prime}\right|^2$. Note that invariants pertaining to the first and second generations generically
involve a single sfermion mass-squared since trilinear couplings do not contribute to their RGEs. For obtaining
RG invariants containing a specific set of parameters it is necessary to form appropriate combinations of these
tabulated ones, as exemplified in the text by a couple of case studies.} \label{table4}
\end{table}

The RGEs for model parameters are all listed in Appendix A. For each quantity the way to MSSM limit is also
described. Similar to the MSSM in Sec. 3, one can construct a number of RG invariants by using the RGEs in
Appendix A. The invariants are tabulated in Table \ref{table4}. The first invariant $I_1^{\prime}$ is nothing but
a direct generalization of the MSSM invariant $I_1$. It expresses the fact that a specific combination of the
inverse gauge coupling-squareds (with arbitrary $c_1$, $c_2$ and $c_3$) is independent of the energy scale.

The U(1)$^{\prime}$ model does not possess an exact RG invariant analogous to $I_2$ in the MSSM. The reason is
that all four Yukawa couplings evolve with scale with their own RG equations; it is not possible form a
scale-invariant combination of the Yukawa-squareds in the absence of a fifth equation that involves the same
couplings (as $d \ln \mu /d t$ does). However, it is still possible to extract some important information about
the UV/IR behaviors of the Yukawa couplings from their RGEs. Indeed, one can show that
\begin{eqnarray}
\label{hseqn} \frac{h_s(Q_2)}{h_s(Q_1)} &=&
  \left(\frac{h_t(Q_1)}{h_t(Q_2)}\right)^{3/7} \left(\frac{h_b(Q_2)}{h_b(Q_1)}\right)^{39/7}
                            \left(\frac{h_{\tau}(Q_1)}{h_{\tau}(Q_2)}\right)^{8/7}\nonumber\\ &\times&
                            \left(\frac{g_3(Q_1)}{g_3(Q_2)}\right)^{64/7} \left(\frac{g_2(Q_2)}{g_2(Q_1)}\right)^{9}
                            \left(\frac{g_1(Q_1)}{g_1(Q_2)}\right)^{5/77}
                            \left(\frac{g_1^{\prime}(Q_2)}{g_1^{\prime}(Q_1)}\right)^{a^{\prime}}\nonumber\\
                            &\times&
                            \exp{\left[- \frac{186}{7} \int_{t_{Q_1}}^{t_{Q_2}} d t^{\prime}\, h_b^2(t^{\prime})\right]}
\end{eqnarray}
with $a^{\prime}$ being a function of the U(1)$^{\prime}$ charges
\begin{eqnarray}
a^{\prime} = \frac{\rho}{42 + 7 \rho \lambda_H} \left( 39 Q_D^2 -8 Q_E^2 + 24 Q_{H_d}^2 - 10 Q_{H_u}^2 - 8 Q_L^2
+36 Q_Q^2 - 7 Q_S^2 -3 Q_U^2\right)
\end{eqnarray}
where $\rho$ and $\lambda_H$ are defined in Appendix A. The importance of (\ref{hseqn}) stems from the fact that
it explicitly expresses $h_s$ in terms of the gauge couplings and rest of the Yukawas. This is important for both
model building and phenomenological purposes since, in general, one has no information about the IR and UV values
of $h_s$ (in contrast to $h_{t,b,\tau}$ whose values at $Q=M_Z$ are known up to the ratio of the doublet VEVs).
The ratio of the IR value of $h_s$ to its UV value depends on all values of $h_b(t)$ in between because of the
integration over $h_b^2$ at the right-hand side. However, this dependence is an extremely weak effect in low
$\tan\beta$ domain where $-(186/7) \int_{t_{Q_1}}^{t_{Q_2}} d t^{\prime}\, h_b^2(t^{\prime}) \sim 10^{-3}$.
Presently, experiments have not shown yet a preferred interval for $\tan\beta$: it can range from 1 to $m_t/m_b$.
However, in U(1)$^{\prime}$ models (and also NMSSM to be discussed in the next section) large (and thus
fine-tuned) $\tan\beta$ regime is not particularly preferred or needed to agree with the LEP bounds
\cite{lep2Zprime}. In fact, as demonstrated in \cite{secluded}, the U(1)$^{\prime}$ models with a secluded sector
naturally realize $\tan\beta \sim {\cal{O}}(1)$ with a heavy enough $Z^{\prime}$ gauge boson. Consequently, the
lesson to be drawn from (\ref{hseqn}) is that given IR and UV values of the gauge and Yukawa couplings then the
ratio $h_s({\rm IR})/h_s({\rm UV})$ is completely determined to an excellent approximation (the validity of which
depends on how small $\tan\beta$ is). This conclusion enables us to introduce an approximate RG invariant
\begin{eqnarray}
I_2^{\prime} \approx h_s\,  h_t^{3/7}\, h_b^{-39/7}\, h_{\tau}^{8/7}\, g_3^{64/7}\, g_2^{-9}\, g_1^{5/77}\,
{g_1^{\prime}}^{-a^{\prime}}
\end{eqnarray}
which exhibits a rather weak scale-dependence especially when the VEVs of the two Higgs doublets are split within
an ${\cal{O}}(1)$ factor. The invariant $I_2^{\prime}$ given in Table \ref{table4} is thus an approximate (albeit
almost exact in low $\tan\beta$ domain) RG invariant.

The U(1)$^{\prime}$ models possess an invariant like $I_3$ in the MSSM (see $I_3^{\prime}$ in Table
\ref{table4}). Indeed, ratio of a gaugino mass to the same gauge group's fine structure constant is an exact RG
invariant. As in the MSSM, such invariants enable one to determine and predict gaugino mass of a given gauge
group when others are given.

The RGEs of the trilinear couplings, given in Appendix A, do not form an exact RG invariant for the reasons valid
for Yukawa couplings. However, one can still establish correlations among the trilinears in order to extract
information about their UV and IR behaviors. For example, the difference between the UV and IR values of $A_s$ is
related to those of the other parameters via
\begin{eqnarray}
A_s(Q_2)-A_s(Q_1) &=& - \frac{372}{7} \int_{t_{Q_1}}^{t_{Q_2}} d t^{\prime}\, h_b^2(t^{\prime})\,
A_b(t^{\prime})\nonumber\\
&+& \frac{3}{7} \left(A_t(Q_1) - A_t(Q_2)\right) + \frac{39}{7} \left( A_b(Q_2) - A_b(Q_1)\right) + \frac{8}{7}
\left(A_{\tau}(Q_1) - A_{\tau}(Q_2)\right)\nonumber\\ &+& \frac{64}{7} M_3(Q_1) \left(
\frac{g_3(Q_2)^2}{g_3(Q_1)^2} -1 \right) - 9 M_2(Q_1) \left( \frac{g_2(Q_2)^2}{g_2(Q_1)^2} -1 \right)\nonumber\\
&+& \frac{5}{77} M_1(Q_1) \left( \frac{g_1(Q_2)^2}{g_1(Q_1)^2} -1 \right) - a^{\prime} M_1^{\prime}(Q_1) \left(
\frac{g_1^{\prime}(Q_2)^2}{g_1^{\prime}(Q_1)^2} -1 \right)
\end{eqnarray}
which depends on all values of $h_b^2 A_b$ in between the UV and IR scales. This dependence, however, is quite
weak in low $\tan\beta$ domain (which is quite natural and does not pose any difficulty with experimental bounds
for the model under concern), and one can safely neglect this contribution. Then, to a good approximation
(validity of which depends on how small $\tan\beta$ is) one can form an RG invariant
\begin{eqnarray}
I_4^{\prime} \approx A_s + \frac{3}{7} A_t - \frac{39}{7} A_b + \frac{8}{7} A_{\tau} + \frac{64}{7} M_3 - 9 M_2 +
\frac{5}{77} M_1 - a^{\prime} M_1^{\prime}
\end{eqnarray}
as is listed in Table \ref{table4}.

The RGEs for scalar soft mass-squareds depend on $S=\mbox{Tr}\left[Y m^2\right]$ and $S^{\prime}=\mbox{Tr}\left[Q
m^2\right]$ whose explicit expression is given in Appendix A. In case soft masses are universal at some scale $Q$
then $S$ vanishes at all scales dues to the absence of hypercharge-graviton-graviton anomaly. In this sense, as
was discussed in detail in Sec. 3 when analyzing the MSSM RGEs, $S$ is a viable probe of universality paradigm.
These properties, however, cannot be continued to $S^{\prime}$ since even if the soft masses are universal
$S^{\prime}$ does not need to vanish because U(1)$^{\prime}$ charges are not guaranteed to cancel the
gravitational anomaly of $Z^{\prime}$ boson. Indeed, the U(1)$^{\prime}$ model is generically anomalous in that
even if gravitational anomaly of U(1)$^{\prime}$ is cancelled there remain all sorts of anomalies (
${\mbox{U(1)}^{\prime}}^{3}$, U(1)$^{\prime}$SU(2)$^{2}$, U(1)$^{\prime}$SU(3)$_c^2$, $\dots$) to be cancelled.
These anomalies cannot be cancelled unless one introduces some exotic matter multiplets which necessarily disrupt
the unification of gauge couplings \cite{cvetic,exotic}. (It is worthy of noting that the model proposed in
\cite{semsi} extends the MSSM with a number of singlet chiral superfields, and determines the singlet
U(1)$^{\prime}$ charges by imposing anomaly cancellation.) Another option, as has recently been pointed out, is
to introduce family non-universal U(1)$^{\prime}$ charges for cancelling anomalies with minimal matter content
\cite{ting}. Both options are beyond the scope of this work which explores RG invariant observables in minimal
U(1)$^{\prime}$ extension of the MSSM. Besides this, scale dependence of $S^{\prime}$ involves all soft masses,
gaugino masses as well as trilinear couplings; it is not as compact as (\ref{seqn}). This continues to be true
unless U(1)$^{\prime}$ charges of opposite-chirality same-flavor fermions obey the same ratios as the
hypercharge. Moreover, U(1)$^{\prime}$ charges of Higgs fields should exhibit a specific proportionality with
their hypercharges. In what follows, we leave aside the question of anomalies and specific representations for
U(1)$^{\prime}$ charges, and simply take $S^{\prime} \equiv 0$ at all scales of interest. (Within specific
U(1)$^{\prime}$ models such as the ones coming from E(6) breaking or family non-universal U(1)$^\prime$ models
the probing power of $S^{\prime}$ can be analyzed explicitly.) With this simplifying assumption the RG invariant
combinations of the soft masses $I_5^{\prime}$--$I_{17}^{\prime}$, in parallel and with respect to those in the
MSSM, are listed in Table \ref{table4}. The modifications in the MSSM invariants are twofold: First of all,  each
invariant picks up an additional contribution proportional to $\left|M_{1}^{\prime}\right|^2$ (there would be an
additional term from $d S^{\prime}/d t$ if $S^{\prime}$ were not taken vanishing). The proportionality constant
involves U(1)$^{\prime}$ beta function and a linear combination of charge-squareds with coefficients identical to
those of the soft mass-squareds relevant for the invariant under consideration. The other modification in MSSM
invariants concerns the presence of Higgs mass-squareds. Indeed, if an MSSM invariant involves $m_{H_{u}}^2$ or
$m_{H_d}^2$ then the corresponding U(1)$^{\prime}$ invariant is necessarily shifted by $-(1/2) m_{S}^2$. The
reason for this is the presence of terms proportional to $h_s^2$ in the beta functions of $m_{H_u}^2$,
$m_{H_d}^2$ and $m_{S}^2$.

A couple of case studies can shed light on certain aspects of the RG invariants in U(1)$^{\prime}$ models. First,
let us consider the RG invariant $I_8^{\prime}$ in Table \ref{table4}. Similar to its MSSM analogue
(\ref{split}), it gives rise to
\begin{eqnarray}
\label{split-U1prime} m_{\widetilde{t}_R}^2(Q_2) + m_{\widetilde{b}_R}^2(Q_2) - 2 m_{\widetilde{t}_L}^2(Q_2) &=& 3 \left|M_2(Q_1)\right|^2
\left[\left(\frac{g_2(Q_2)}{g_2(Q_1)}\right)^4 -1\right]\nonumber\\ &-&  \frac{1}{11} \left|M_1(Q_1)\right|^2
\left[\left(\frac{g_1(Q_2)}{g_1(Q_1)}\right)^4 -1\right] \nonumber\\
&+&\frac{\rho}{6 +  \rho \lambda_H} (Q_{U}^2 + Q_D^2 - 2 Q_Q^2)\nonumber\\ &\times&
\left|M_1^{\prime}(Q_1)\right|^2 \left[\left(\frac{g_1^{\prime}(Q_2)}{g_1^{\prime}(Q_1)}\right)^4 -1\right]
\end{eqnarray}
when soft mass-squareds are all universal at some scale $Q=Q_1$. This relation is independent of the Higgs sector
parameters; it is sensitive to only the isospin and Abelian group factors. In fact, it feels whether the gauge
sector is minimal or not by  the inclusion of the corresponding gaugino mass in the sum rule. Therefore, via the
last term $\propto \left|M_1^{\prime}\right|^{2}$, it obtains the potential of probing the existence of an
additional U(1)$^{\prime}$ gauge invariance provided that one can perform precise measurements and consistency
checks with other sectors of the theory. Of course, (\ref{split-U1prime}) can be used to establish a relation
between the stop and sbottom mixing angles in the same spirit as (\ref{angles}).

One notices that it is not possible to construct an RG invariant which feels only the extensions in the Higgs
sector. The reason is that in a given sum rule each mass-squared parameter is accompanied by an additional term
$\propto Q^2 \left|M_1^{\prime}\right|^2$ any attempt at cancelling terms involving $\left|M_1^{\prime}\right|^2$
necessarily ends up with cancelling $m_S^2$ contribution. Of course, within a specific representation for
U(1)$^{\prime}$ invariance charges of various fields could be correlated to cancel out without nullifying the
coefficient of $m_S^2$ in the final sum rule. It was this property of hypercharge symmetry that allowed us to
arrive at (\ref{higgses}) in the MSSM section above. For instance, if $Q_{E}^2$ happens to be proportional to
$Q_{H_u}^2 + Q_{H_d}^2 - Q_S^2 - 3 Q_Q^2 - Q_L^2$ then $I_{14}^{\prime}$ and $I_{13}^{\prime}$ can be used to
relate mass-squareds of Higgs fields to those of sfermions and MSSM gauginos.

There exist certain RG-invariant combinations of the soft-mass squareds which depend on extensions in neither the
gauge nor the Higgs sectors. Invariants of this kind can be easily constructed by linearly combining those in
Table \ref{table4}. For instance,
\begin{eqnarray}
\label{combine}
I_8^{\prime} + 2 I_9^{\prime} - I_{10}^{\prime} - I_{11}^{\prime} &=& \left( m_{\widetilde{t}_R}^2-m_{\widetilde{u}_R}^2\right) +
\left(m_{\widetilde{b}_R}^2-m_{\widetilde{d}_R}^2\right) - 2 \left(m_{\widetilde{t}_L}^2 -m_{\widetilde{u}_L}^2\right)
\end{eqnarray}
is an RG invariant in both MSSM and U(1)$^{\prime}$ models. Clearly, this kind of quantities are completely
insensitive to modifications in the Higgs and gauge sectors; they exclusively probe the sfermion sector.

In general, within specific string or supergravity models, the soft parameters exhibit various interrelations
which give rise to a spectrum of discriminative values for the RG invariants. This can be used for predicting
what specific model could be responsible for supersymmetry breaking. Indeed, as one recalls from discussions of
the MSSM invariants within no-scale supergravity, dilaton-domination and flux-induced soft terms, measurements of
RG invariants could be a useful tool for determining which high scale model is operating as more and more data
accumulate. For the U(1)$^{\prime}$ model under concern it is convenient to discuss the predictive powers of RG
invariants with respect to the MSSM ones tabulated in Tables \ref{table2} and \ref{table3}: First of all, if the
UV scale model realizes a universal gaugino mass $M$ then each of $I_{5}^{\prime}/M^2$--$I_{17}^{\prime}/M^2$ is
shifted by an amount
\begin{eqnarray}
\frac{\rho}{6 +  \rho \lambda_H} \left(a_X Q_{X}^2 + b_Y Q_Y^2 + c_Z Q_Z^2+ \dots\right)
\end{eqnarray}
if the invariant under concern is composed of $a_X m_X^2 + b_Y m_Y^2 + c_Z m_Z^2 + \dots$ If an invariant does
not contain Higgs soft mass-squareds or $m_0= 0$ for the model under concern then this is the only modification
in an RG invariant with respect to its MSSM value given in Tables \ref{table2} and \ref{table3}. On the other
hand, if an invariant consists of the Higgs masses, $e.g.$ $I_6^{\prime}$, $I_7^{\prime}$, $I_{14}^{\prime}$,
then departure from the MSSM expression occurs in both $m_0^2$ and $M^2$ directions. The discriminative power of
an RG invariant depends on its correlation with others for a given GUT-scale configuration. Generically, if the
mass-squareds of Higgs doublets are present in an invariant so is that of the Higgs singlet.

The RG invariants do have interesting implications also for chargino and neutralino sectors of the model. For
instance, from the product of two chargino masses one immediately finds that the quantity
\begin{eqnarray}
\label{char-U1prime} \frac{1}{\langle S \rangle} \left(M_{\chi_1^\pm} M_{\chi_2^\pm} + M_W^2 \sin 2 \beta\right)
\left( h_t^{3/7}\, h_b^{-39/7}\, h_{\tau}^{8/7}\, g_3^{64/7}\, g_2^{-11}\, g_1^{5/77}\,
{g_1^{\prime}}^{-a^{\prime}}\right)
\end{eqnarray}
is an approximate RG invariant. This RG invariant differs from its MSSM analogue (\ref{char}) by modifications in
powers of the gauge and Yukawa couplings and by the presence of the singlet VEV $\langle S \rangle$. The presence
of the singlet VEV stems from the fact that the $\mu$ parameter in the MSSM is generated dynamically by the
singlet VEV: $\mu_{eff} = h_s \langle S \rangle$.

The neutralino sector is sensitive to both the Higgs singlet and U(1)$^{\prime}$ gaugino. First of all, sum of
the neutralino masses obey
\begin{eqnarray}
\label{nsum-U1prime} \sum_{i=1}^{6} M_{\chi_i^0} = M_1^{\prime} + M_1 + M_2 = \left( {g_1^{\prime}}^2 + g_1^2 +
g_2^2\right) \frac{M_{1/2}}{g_0^2}
\end{eqnarray}
when the gaugino masses unify into $M_{1/2}$ at the scale where gauge couplings do into $g_0$. The sum of the
squared-masses of neutralinos depend on both $M_1^{\prime}$ and $h_s \langle S \rangle$. Therefore, a correlated
analysis of neutralino and chargino sectors provide important information on whether the MSSM is extended by new
gauge symmetries and/or new Higgs representations. The neutralino sector admits several sum/product rules similar
to (\ref{carpim}) in the MSSM, and they can be used to form novel RG invariant combinations of the
chargino/neutralino parameters in the same spirit as (\ref{nsum-U1prime}) and (\ref{char-U1prime}). One keeps in
mind, however, that invariants involving the Higgs singlet is always approximate in the sense of (\ref{hseqn}).

In this section we have derived a number of RG invariants in U(1)$^{\prime}$ models, and performed certain case
studies for highlighting their phenomenological relevance. These invariants (albeit approximate for Yukawa
couplings and trilinear soft terms) could be useful for establishing gauge and/or Higgs extension with respect to
the MSSM.

\section{The RG Invariants in the NMSSM}
The next-to-minimal supersymmetric model possesses no gauge extension with respect to MSSM. Its Higgs sector is
spanned by $\widehat{H}_u$, $\widehat{H}_d$ and $\widehat{S}$ so that
\begin{eqnarray}
\label{rigid-Higgs-NMSSM} \widehat{W}_{Higgs} = h_s \widehat{S} \widehat{H}_u \widehat{H}_d + \frac{k_s}{6}
\widehat{S}^3
\end{eqnarray}
and
\begin{eqnarray}
 \label{soft-Higgs-NMSSM}
-{\cal{L}}_{soft}^{Higgs}&=& m_{H_u}^2 H_u^{\dagger} H_u + m_{H_d}^2 H_d^{\dagger} H_d + m_S^2 S^{\dagger} S +
\left[ h_s A_s S {H}_u\cdot {H}_d + \frac{k_s}{6} A_k S^3 + \mbox{h.c.}\right],\nonumber\\
-{\cal{L}}_{soft}^{gaugino}&=& \frac{1}{2}\sum_{a=3,2,1} \left[ M_{a} \lambda_{a} \lambda_{a} +
\mbox{h.c.}\right]
\end{eqnarray}
where the singlet cubic interaction in the superpotential is needed to generate a potential for $S$ ( this field
does not have a D-term support to obtain a potential). The induction of effective $\mu$ and $B$ parameters are
similar to those of the U(1)$^{\prime}$ model given in (\ref{effectives}). The main difference from the
U(1)$^{\prime}$ model lies in the fact that the $\widehat{S}$ is a pure singlet (in both MSSM and NMSSM) so that
it is allowed to develop a cubic interaction in the superpotential.

The RGEs of the rigid and soft parameters of the model are all listed in Appendix B. We also discuss the MSSM
limits of individual RGEs for easy comparison of the corresponding RG invariants. The RG-invariant quantities in the model
are listed in Table \ref{table5}. Obviously, the RG invariant combinations of gauge couplings, $I_1^{\prime\prime}$,
remain the same as in the MSSM.

\begin{table}[h]
\begin{center}
\begin{tabular}{|c||c|}
  \hline
  Number  & RG Invariant \\ \hline \hline
  $I_1^{\prime \prime}$  & $\frac{c_1}{g_1^2}+\frac{c_2}{g_2^2}+\frac{33 c_1 +5 c_2}{15 g_3^2}$  \\  \hline
  $I_2^{\prime \prime}$  & $\approx k_s\,  h_t^{3/7}\, h_b^{15/7}\, h_{\tau}^{-6/7}\, h_s^{-3}\, g_3^{-832/189}\,
g_2^{-27/7}\, g_1^{-23/77}$ \\ \hline\hline $I_3^{\prime \prime}$  & $\frac{M_{a}}{g_{a}^2}\; (a=1,2,3) $ \\
\hline $I_4^{\prime \prime}$  & $\approx A_k + \frac{3}{7} A_t + \frac{15}{7} A_b - \frac{6}{7} A_{\tau} -
\frac{382}{189} M_3 - \frac{27}{7} M_2 - \frac{23}{77} M_1$ \\ \hline\hline $I_5^{\prime \prime}$  & $ I_5$\\
\hline $I_8^{\prime \prime}$  & $ I_8 $\\\hline $I_9^{\prime}$  & $I_9$\\ \hline $I_{10}^{\prime \prime}$  &
$I_{10}$\\ \hline $I_{11}^{\prime \prime}$  & $ I_{11}$\\ \hline $I_{12}^{\prime \prime}$  & $ I_{12}$ \\ \hline
$I_{13}^{\prime \prime}$  & $ I_{13}$\\ \hline  $I_{16}^{\prime \prime}$  & $I_{16}$\\ \hline $I_{17}^{\prime
\prime}$ & $I_{17}$\\\hline
\end{tabular}
\end{center}
\caption{The RG invariant combinations of rigid and soft parameters in the NMSSM ( $c_{1}$ and $c_2$ are
arbitrary constants). Here $I_2^{\prime}$ and $I_4^{\prime}$ are approximate invariants derived in the text. The
invariants constructed from scalar mass-squareds are written in terms of the MSSM invariants in Table
\ref{table1}. The missing rows (with respect to Table \ref{table1}) indicate that there are no analogous RG
invariant combinations of the scalar soft mass-squareds (the ones that depend on the Higgs sector parameters).}
\label{table5}
\end{table}

In close similarity to U(1)$^{\prime}$ models, the Yukawa couplings do not possess an exact RG invariant.
However, it is still possible to express one of the Yukawas in terms of the rest and gauge couplings. For
instance, the singlet cubic coupling is related to others via
\begin{eqnarray}
\label{kseqn} \frac{k_s(Q_2)}{k_s(Q_1)} &=&
  \left(\frac{h_t(Q_1)}{h_t(Q_2)}\right)^{3/7} \left(\frac{h_b(Q_1)}{h_b(Q_2)}\right)^{15/7}
                            \left(\frac{h_{\tau}(Q_2)}{h_{\tau}(Q_1)}\right)^{6/7}
                            \left(\frac{h_{s}(Q_2)}{h_{s}(Q_1)}\right)^{3}
                            \nonumber\\ &\times&
                            \left(\frac{g_3(Q_2)}{g_3(Q_1)}\right)^{832/189} \left(\frac{g_2(Q_2)}{g_2(Q_1)}\right)^{27/7}
                            \left(\frac{g_1(Q_2)}{g_1(Q_1)}\right)^{23/77}\nonumber\\
                            &\times&
                            \exp{\left[\frac{197}{7} \int_{t_{Q_1}}^{t_{Q_2}} d t^{\prime}\, h_b^2(t^{\prime})\right]}
\end{eqnarray}
The importance of this relation stems from the fact that it explicitly expresses $k_s$ in terms of the gauge
couplings and rest of the Yukawas. This is important for both model building and phenomenological purposes since,
in general, one has no information about the IR and UV values of both $h_s$ and $k_s$ (in contrast to
$h_{t,b,\tau}$ whose values at $Q=M_Z$ are known up to the ratio of the doublet VEVs), and it is advantageous to
know at least one's value in terms of the rest. The ratio of the IR value of $k_s$ to its UV value depends on all
values of $h_b(t)$ in between because of the integration over $h_b^2$ at the right-hand side. However, this
dependence is an extremely weak effect in low $\tan\beta$ domain where $(197/7) \int_{t_{Q_1}}^{t_{Q_2}} d
t^{\prime}\, h_b^2(t^{\prime}) \sim 10^{-3}$. Presently, experiments have not shown yet a preferred interval for
$\tan\beta$: it can range from 1 to $m_t/m_b$. However, in NMSSM large (and thus fine-tuned) $\tan\beta$ regime
is not particularly preferred or needed to explain the LEP limits \cite{LEP2-nmssm}. Consequently, (\ref{kseqn}) implies that,
given IR and UV values of the gauge and Yukawa couplings, then the ratio $k_s({\rm IR})/k_s({\rm UV})$ is
completely determined to an excellent approximation (the validity of which depends on how small $\tan\beta$ is).
This conclusion enables us to introduce an approximate RG invariant
\begin{eqnarray}
I_2^{\prime \prime} \approx k_s\,  h_t^{3/7}\, h_b^{15/7}\, h_{\tau}^{-6/7}\, h_s^{-3}\, g_3^{-832/189}\,
g_2^{-27/7}\, g_1^{-23/77}
\end{eqnarray}
which exhibits a rather weak scale-dependence especially when the VEVs of the two Higgs doublets are split within
an ${\cal{O}}(1)$ factor \cite{LEP2-nmssm}. The invariant $I_2^{\prime \prime}$ given in Table \ref{table5} is thus an approximate
RG invariant.

The ratio of the gaugino masses to the corresponding fine structure constant, $I_3^{\prime\prime}$ in Table
\ref{table5}, is an RG invariant, and it equals the corresponding invariant in the MSSM.

The behaviors of the trilinear couplings are similar to Yukawas. They do not admit an exact RG invariant.
However, one can correlate their UV and IR values as in the U(1)$^{\prime}$ models. For example, the difference
between the UV and IR values of $A_k$ is related to those of the other parameters via
\begin{eqnarray}
A_k(Q_2)-A_k(Q_1) &=& \frac{384}{7} \int_{t_{Q_1}}^{t_{Q_2}} d t^{\prime}\, h_b^2(t^{\prime})\,
A_b(t^{\prime})\nonumber\\
&+& \frac{3}{7} \left(A_t(Q_1) - A_t(Q_2)\right) + \frac{15}{7} \left( A_b(Q_1) - A_b(Q_2)\right)\nonumber\\ &+&
\frac{6}{7} \left(A_{\tau}(Q_2) - A_{\tau}(Q_1)\right) + 3 \left(A_{s}(Q_2) - A_{s}(Q_1)\right)\nonumber\\ &-&
\frac{382}{189} M_3(Q_1) \left(
\frac{g_3(Q_2)^2}{g_3(Q_1)^2} -1 \right) - \frac{27}{7} M_2(Q_1) \left( \frac{g_2(Q_2)^2}{g_2(Q_1)^2} -1 \right)\nonumber\\
&-& \frac{23}{77} M_1(Q_1) \left( \frac{g_1(Q_2)^2}{g_1(Q_1)^2} -1 \right)
\end{eqnarray}
which depends on all values of $h_b^2 A_b$ in between the UV and IR scales. This dependence, however, is quite
weak in low $\tan\beta$ domain, and as experiments are not pushing for high $\tan\beta$ regime for the NMSSM,
this dependence on  $h_b^2 A_b$ can safely be neglected. Then, in low $\tan\beta$ regime, one can form an
approximate RG invariant
\begin{eqnarray}
I_4^{\prime} \approx A_k + \frac{3}{7} A_t + \frac{15}{7} A_b - \frac{6}{7} A_{\tau} - \frac{382}{189} M_3 -
\frac{27}{7} M_2 - \frac{23}{77} M_1
\end{eqnarray}
as listed in Table \ref{table5}.

As in the MSSM and U(1)$^{\prime}$ models the soft squared-mass parameters do also form a number of RG
invariants. These are listed in Table \ref{table5}. Perhaps, the most interesting aspect of the NMSSM is that its
Higgs sector parameters do not admit any RG invariant. The reason is that the RG running of $m_S^2$ is
necessarily affected by the cubic singlet coupling via $\left( 3 m_S^2 + \left|A_k\right|^2\right) k_s^2$ whereas
running of the squared-masses of other fields do not involve terms $\propto k_s^2$. Hence, this term cannot be
cancelled to form an invariant, and therefore, the MSSM RG invariants $I_6$, $I_7$, $I_{14}$, and $I_{15}$ (which
consist of the Higgs squared-masses) in Table \ref{table1} do not possess any analogue in Table \ref{table5}.
Physically, this is related to the fact that neither $F$ terms nor soft terms generate operators of the form
$k_s^2 \left|S\right|^2 \left( \left|H_{u}\right|^2, \left|H_d\right|^2, \left|\widetilde{Q}\right|^2, \dots
\right)$. It is convenient to dwell on this point by examining one of the would-be invariants. For instance, in the
present model $I_{14}$ in Table \ref{table1} generalizes to
\begin{eqnarray}
\label{noninv} \frac{d}{d t} \left(I_{14} - m_{S}^2\right) = - \left( 3 m_S^2 + \left|A_k\right|^2 \right) k_s^2
\end{eqnarray}
so that $I_{14} -m_S^2$ is not a scale-invariant observable; it exhibits a nontrivial RG flow unless ($i$) $k_s =
0$ or ($ii$) $m_S^2 = - \left|A_k\right|^2/3$. The former is disfavored for it gives rise to a flat direction for
$S$ \cite{nmssm}. The latter, however, represents a fixed point solution for $m_{S}^2$ in that at the scale it holds $m_S^2$
is guaranteed to be negative and hence the theory below $\left|A_k\right|$ generates the MSSM as an effective
theory. Clearly, if $\left|A_k\right| \sim {\cal{O}}({\rm TeV})$ the MSSM Higgs sector gets correctly stabilized
at the desired scale.

Looking from a different angle, (\ref{noninv}) provides an experimental testing ground (presumably after LHC
$\oplus$ ILC) for knowing if the model under concern is NMSSM or U(1)$^{\prime}$ extension of the MSSM. Indeed,
in U(1)$^{\prime}$ models the right hand side of (\ref{noninv}) is  $\propto {g_1^{\prime}}^2
\left|M_1^{\prime}\right|^2$ and it can be written as a total derivative to form the invariant $I_{14}^{\prime}$
in Table \ref{table4}. Moreover, as depicted in Table \ref{table4} all RG invariants of soft mass-squareds
systematically contain $\left|M_1^{\prime}\right|^2$ so that after sufficient number of precise measurements one
can make sure if the model under concern involves a new gaugino or not. In contrast to this, the right hand side
of (\ref{noninv}) cannot be written as a total derivative; moreover, it shows up only in those would-be invariants
which include the Higgs soft-mass squareds. The rest of the invariants, as shown in Table \ref{table5}, are
identical to those in the MSSM. In this sense, non-invariance of the Higgs sector parameters can provide a viable
signal of NMSSM in future collider tests.

The RG-invariant combinations of the squared soft masses listed in Table \ref{table5} give rise to certain
correlations or sum rules which are identical to those derived in the MSSM. For instance, $I_{8}^{\prime \prime}$
relates stop plus sbottom splittings to the isospin and hypercharge gaugino masses in the same way as
(\ref{split}).

The RG invariant combinations of the chargino/neutralino systems are similar to ones in U(1)$^{\prime}$ models.
Indeed, (\ref{char-U1prime}) now becomes
\begin{eqnarray}
\label{char-NMSSM} \frac{1}{\langle S \rangle} \left(M_{\chi_1^\pm} M_{\chi_2^\pm} + M_W^2 \sin 2 \beta\right)
\left( h_t^{-1/7}\, h_b^{-5/7}\, h_{\tau}^{2/7}\, g_3^{832/567}\, g_2^{-5/7}\, g_1^{23/231}\right)
\end{eqnarray}
is an approximate RG invariant in the sense of (\ref{kseqn}). The singlet VEV $\langle S \rangle$ arises due to
the dynamical origin of the MSSM $\mu$ parameter: $\mu_{eff} = h_s \langle S \rangle$.

The neutralino sector is interesting in that sum of the neutralino masses satisfy
\begin{eqnarray}
\label{nsum-NMSSM} \sum_{i=1}^{5} M_{\chi_i^0} = M_1 + M_2 = \left( g_1^2 + g_2^2\right) \frac{M_{1/2}}{g_0^2}
\end{eqnarray}
which is identical to the MSSM prediction. The NMSSM effects show up when we consider sum of the neutralino
mass-squareds or when we consider their products. Such quantities, too, can be expressed in terms of the RG
invariants at low values of $\tan\beta$. Their validity and construction are not different than
(\ref{char-NMSSM}).

In this section we have analyzed the RGEs of the NMSSM for determining RG invariant combinations of the
lagrangian parameters. Concerning the scale dependencies of the Yukawa couplings and trilinear soft terms, the
behavior is similar to U(1)$^{\prime}$ model. On the other hand, RG invariants made up of gauge couplings and
scalar soft mass-squareds are the same as in the MSSM. The model radically differs from the MSSM and
U(1)$^{\prime}$ model due to the absence of RG invariants containing the Higgs mass-squareds.

\section{Conclusion}

In this work, using one loop RGEs, we have derived a number of scale-invariant observables
in softly-broken supersymmetric models, and illustrated their
phenomenological implications by various case studies. We have first
studied the MSSM and then its minimal extensions, U(1)$^{\prime}$ models and NMSSM, in
a comparative manner.

In general, each supersymmetric model possesses RG invariants in gauge, Yukawa and
soft-breaking sectors. The invariants of the MSSM, of U(1)$^{\prime}$ model and
of the NMSSM are listed in Tables \ref{table1}, \ref{table4} and \ref{table5},
respectively. In general, RG invariants vary from model to model though
those associated exclusively with their common part, the sfermion sector,
may be combined to obtain invariants valid for all three models (see $e.g.$
the combination \ref{combine}).

The RG-invariant combinations of the gauge couplings and gaugino masses
are idetical for the MSSM and NMSSM whereas additional gauge coupling
and mass of the associated gaugino introduces additional terms for the U(1)$^{\prime}$
model. For phenomenological purposes, such invariants prove particularly useful
when gauge couplings and gaugino masses unify at high scale.

The $\mu$ parameter, Yukawa couplings and gauge couplings combine to form an RG
invariant in the MSSM. This, however, is not the case in U(1)$^{\prime}$ models
and NMSSM. In these models, the best one can do is to correlate one of the Yukawas
in terms of the rest so that an approximate RG invariant emerges within a specific
domain of the parameter space. In fact, $I_2^{\prime}$ in Table \ref{table4} and
$I_2^{\prime \prime}$ in Table \ref{table5} serve as RG invariants only for
low values of $\tan\beta$.

The Higgs bilinear soft mass $B$ and sfermion-sfermion-Higgs trilinear couplings
form an exact RG invariant in the MSSM. However, for U(1)$^{\prime}$ models
and NMSSM there are no such invariants, and as for the Yukawa sector, all one
can do is to realize an approximate invariant in a specific domain of the
parameter space. In fact, $I_4^{\prime}$ in Table \ref{table4} and
$I_4^{\prime \prime}$ in Table \ref{table5} behave as RG invariants only at
low values of $\tan\beta$.

The MSSM possesses a number of RG invariants containing the squared-masses
of the scalars. They can be grouped into two classes: The ones that are
not sensitive to whether the scalar masses attain a universal configuration
and the ones that are sensitive (via the quantity $S$) to such a configuration.
Therefore, the $S$ dependence of the invariants serves as a tool for probing
the UV scale correlations of the soft mass-squareds (as part of the minimal
supergravity configuration). Moreover, as shown in Tables \ref{table3} and \ref{table4},
the invariants take on a specific set of values for each mechanism of supersymmetry
breaking, and therefore, they can be used for determining the origin of supersymmetry
breaking.

The RG invariants of scalar mass-squareds in the MSSM get modified by
the U(1)$^{\prime}$ gaugino mass and by the singlet mass-squared. In particular,
those MSSM invariants which depend on the Higgs mass-squareds are generically
generically shifted by the singlet mass-squared. It is possible to form
new invariants that involve only the U(1)$^{\prime}$ gaugino mass. On the other hand,
invariants that depend only on the singlet mass-squared cannot be formed (unless
one uses a specific representation for U(1)$^{\prime}$ charges).

The situation in the NMSSM is interesting in that the Higgs mass-squareds cannot
be combined to form an invariant because of the presence of cubic singlet coupling
in the superpotential. This non-invariance itself can be useful for model identification
at future collider studies. On the other hand, mass-squareds of scalar quarks and
leptons admit RG-invariant configurations that are identical to those in the MSSM.

In general, the RG invariants are useful for both model-building and
phenomenological purposes as they make various indirect relations
manifest. This enhances one's knowledge of various dependencies and
correlations among the model parameters. Moreover, they give rise to
certain sum rules which can be quite useful for determining the underlying
model and origin of supersymmetry breaking as data accumulate at detectors.
Various relations require an accurate measurement of a subset of parameters
which could be possible after LHC$\oplus$ILC.

\section{Acknowledgements}
The author thanks Lisa Everett for discussions on RGEs. This
work was partially supported by Turkish Academy of
Sciences through GEBIP grant, and by the Scientific and
Technical Research Council of Turkey through project 104T503.

\newpage
\section*{Appendix A. Renormalization Group Equations in U(1)$^\prime$ Model}
\setcounter{equation}{0}
\def\theequation{A.\arabic{equation}}
In this Appendix we list down the RGEs for U(1)$^{\prime}$ models by extending \cite{cvetic} to cases with
finite bottom and tau Yukawas in a way including all three generations of sfermions. The one-loop RGEs of the
gauge couplings are given by
\begin{eqnarray}
\frac{d g_3}{d t} &=& \left( 2 N_F - 9\right)g_3^3\nonumber\\
\frac{d g_2}{d t} &=& \left( 2 N_F - 5\right)g_2^3\nonumber\\
\frac{d g_1}{d t} &=& \left( 2 N_F + \frac{3}{5}\right)g_1^3\nonumber\\
\frac{d g_1^{\prime}}{d t} &=& \left( 2 N_F + \rho \lambda_H\right){g_1^{\prime}}^3
\end{eqnarray}
where $t\equiv (4 \pi)^{-2}\, \ln Q/M_{GUT}$, $N_F=3$, $\lambda_H = Q_{H_d}^2 + Q_{H_u}^2 + \frac{1}{2} Q_S^2$,
and
\begin{eqnarray}
\rho = \frac{4}{6 Q_Q^2 + 3\left(Q_U^2+Q_D^2\right) + 2 Q_L^2 + Q_E^2}
\end{eqnarray}
which is obtained by requiring $g^2_a \mbox{Tr}\left[Q^2\right]$ to be identical for all group factors. The
U(1)$^{\prime}$ charges $Q_{H_u,\dots,E}$ are family-universal. The corresponding MSSM RGEs are recovered by
setting $g_1^{\prime}=0.$

The evolutions of the superpotential parameters are given by
\begin{eqnarray}
\frac{d h_t}{d t} &=& h_t \left( 6 h_t^2 + h_b^2 + h_s^2 - \frac{16}{3} g_3^2 - {3} g_2^2 - \frac{13}{15}
g_1^2 - \rho \left( Q_{H_u}^2 + Q_{Q}^2 + Q_U^2\right) {g_1^{\prime}}^2\right)\nonumber\\
\frac{d h_b}{d t} &=& h_b \left( 6 h_b^2 + h_t^2 + h_{\tau}^2 + h_s^2 - \frac{16}{3} g_3^2 - {3} g_2^2 -
\frac{7}{15}
g_1^2 - \rho \left( Q_{H_d}^2 + Q_{Q}^2 + Q_D^2\right) {g_1^{\prime}}^2\right)\nonumber\\
\frac{d h_{\tau}}{d t} &=& h_{\tau} \left( 4 h_{\tau}^2 + 3 h_b^2 + h_s^2 - {3} g_2^2 - \frac{9}{5}
g_1^2 - \rho \left( Q_{H_d}^2 + Q_{L}^2 + Q_E^2\right) {g_1^{\prime}}^2\right)\nonumber\\
\frac{d h_s}{d t} &=& h_s \left( 4 h_s^2 + 3 h_t^2 + 3 h_b^2 + h_{\tau}^2 - 3 g_2^2 - \frac{3}{5} g_1^2 - \rho
\left( Q_{H_d}^2 + Q_{H_u}^2 + Q_S^2\right) {g_1^{\prime}}^2\right)
\end{eqnarray}
which reduce to the corresponding RGEs in the MSSM after setting $g_{1}^{\prime} = 0$, identifying $d \ln h_s
/dt$ with  $d \ln \mu / dt$ in the last equation, and taking $h_s =0$ everywhere else (since now a dynamical
field $\widehat{S}$ does not exist at all).

The gaugino masses evolve as
\begin{eqnarray}
\frac{d M_3}{d t} &=& \left( 4 N_F - 18\right)g_3^2 M_3\nonumber\\
\frac{d M_2}{d t} &=& \left( 4 N_F - 10\right)g_2^2 M_2\nonumber\\
\frac{d M_1}{d t} &=& \left( 4 N_F + \frac{6}{5}\right)g_1^2 M_1\nonumber\\
\frac{d M_1^{\prime}}{d t} &=& \left( 4 N_F +\rho \left( 2 Q_{H_d}^2 + 2 Q_{H_u}^2 +
Q_S^2\right)\right){g_1^{\prime}}^2 M_1^{\prime}
\end{eqnarray}
which reduce to the RGEs in the MSSM after setting $M_1^{\prime} = 0$.

The RG evolutions of the trilinear couplings are given by
\begin{eqnarray}
\frac{d A_t}{d t} &=& 2 \left( 6 h_t^2 A_t + h_b^2 A_b + h_s^2 A_s\right)\nonumber\\ &+&2 \left( \frac{16}{3}
g_3^2 M_3 + {3} g_2^2 M_2 + \frac{13}{15} g_1^2 M_1 + \rho
\left( Q_{H_u}^2 + Q_{Q}^2 + Q_U^2\right) {g_1^{\prime}}^2 M_1^{\prime}\right)\nonumber\\
\frac{d A_b}{d t} &=& 2 \left( 6 h_b^2 A_b + h_t^2 A_t + h_{\tau}^2 A_{\tau} + h_s^2 A_s\right)\nonumber\\ &+&2
\left( \frac{16}{3} g_3^2 M_3 + {3} g_2^2 M_2 + \frac{7}{15}
g_1^2 M_1 + \rho \left( Q_{H_d}^2 + Q_{Q}^2 + Q_D^2\right) {g_1^{\prime}}^2 M_1^{\prime}\right)\nonumber\\
\frac{d A_{\tau}}{d t} &=& 2 \left( 4 h_{\tau}^2 A_{\tau} + 3 h_b^2 A_{b} + h_s^2 A_s \right)\nonumber\\ &+& 2
\left( {3} g_2^2 M_2 + \frac{9}{5}
g_1^2 M_1 + \rho \left( Q_{H_d}^2 + Q_{L}^2 + Q_E^2\right) {g_1^{\prime}}^2 M_1^{\prime}\right)\nonumber\\
\frac{d A_s}{d t} &=& 2 \left( 4 h_s^2 A_s + 3 h_t^2 A_t + 3 h_b^2 A_b + h_{\tau}^2 A_{\tau} \right)\nonumber\\
&+& 2 \left( 3 g_2^2 M_2 + \frac{3}{5} g_1^2 M_1 + \rho \left( Q_{H_d}^2 + Q_{H_u}^2 + Q_S^2\right)
{g_1^{\prime}}^2 M_1^{\prime}\right)
\end{eqnarray}
where RGEs of the corresponding quantities within the MSSM are obtained by $g_1^{\prime} = 0$, $h_s =0$ and $A_s
= B$, $B$ being the Higgs soft bilinear coupling.

The scalar soft mass--squared parameters evolve according to
\begin{eqnarray}
\frac{d m_{H_u}^2}{d t} &=& 2 \left(m_{H_u}^2 + m_{H_d}^2 + m_{S}^2 + \left|A_s\right|^2\right) h_s^2\nonumber\\
                        &+& 6 \left(m_{H_u}^2 + m_{\widetilde{t}_L}^2 + m_{\widetilde{t}_R}^2 + \left|A_t\right|^2\right) h_t^2 \nonumber\\
                        &-& 8 \left( \frac{3}{4} g_2^2 \left|M_2\right|^2 + \frac{3}{20} g_1^2
                        \left|M_1\right|^2 + \frac{1}{2} \rho Q_{H_u}^2 {g_1^{\prime}}^2
                        \left|M_1^{\prime}\right|^2\right)\nonumber\\
                        &+& \frac{3}{5} g_1^2 S + \rho Q_{H_u} {g_1^{\prime}}^2 S^{\prime}\nonumber\\
\frac{d m_{H_d}^2}{d t} &=& 2 \left(m_{H_u}^2 + m_{H_d}^2 + m_{S}^2 + \left|A_s\right|^2\right) h_s^2\nonumber\\
                        &+& 2 \left(m_{H_d}^2 + m_{\widetilde{\tau}_L}^2 + m_{\widetilde{\tau}_R}^2 + \left|A_{\tau}\right|^2\right)
h_{\tau}^2\nonumber\\
                        &+& 6 \left(m_{H_d}^2 + m_{\widetilde{t}_L}^2 + m_{\widetilde{b}_R}^2 + \left|A_b\right|^2\right) h_b^2 \nonumber\\
                        &-& 8 \left( \frac{3}{4} g_2^2 \left|M_2\right|^2 + \frac{3}{20} g_1^2
                        \left|M_1\right|^2 + \frac{1}{2} \rho Q_{H_d}^2 {g_1^{\prime}}^2
                        \left|M_1^{\prime}\right|^2\right)\nonumber\\
                        &-& \frac{3}{5} g_1^2 S + \rho Q_{H_d} {g_1^{\prime}}^2 S^{\prime}\nonumber\\
\frac{d m_{S}^2}{d t} &=& 4 \left(m_{H_u}^2 + m_{H_d}^2 + m_{S}^2 + \left|A_s\right|^2\right) h_s^2\nonumber\\
                         &-& 4 \rho Q_{S}^2 {g_1^{\prime}}^2
                        \left|M_1^{\prime}\right|^2
                        + \rho Q_{S} {g_1^{\prime}}^2 S^{\prime}\nonumber\\
\frac{d m_{\widetilde{t}_L}^2}{d t} &=& 2 \left(m_{\widetilde{t}_L}^2 + m_{H_d}^2 + m_{\widetilde{b}_R}^2 + \left|A_b\right|^2\right)
h_b^2\nonumber\\
                        &+&2 \left(m_{\widetilde{t}_L}^2 + m_{H_u}^2 + m_{\widetilde{t}_R}^2 + \left|A_t\right|^2\right) h_t^2\nonumber\\
                        &-& 8 \left( \frac{4}{3} g_3^2 \left|M_3\right|^2 + \frac{3}{4} g_2^2 \left|M_2\right|^2 +
                        \frac{1}{60} g_1^2 \left|M_1\right|^2 + \frac{1}{2} \rho Q_{Q}^2 {g_1^{\prime}}^2
                        \left|M_1^{\prime}\right|^2\right)\nonumber\\
                        &+& \frac{1}{5} g_1^2 S + \rho Q_{Q} {g_1^{\prime}}^2 S^{\prime}\nonumber\\
\frac{d m_{\widetilde{t}_R}^2}{d t} &=& 4 \left(m_{\widetilde{t}_L}^2 + m_{H_u}^2 + m_{\widetilde{t}_R}^2 + \left|A_t\right|^2\right)
h_t^2\nonumber\\
                        &-& 8 \left( \frac{4}{3} g_3^2 \left|M_3\right|^2 +
                        \frac{4}{15} g_1^2 \left|M_1\right|^2 + \frac{1}{2} \rho Q_{U}^2 {g_1^{\prime}}^2
                        \left|M_1^{\prime}\right|^2\right)\nonumber\\
                        &-& \frac{4}{5} g_1^2 S + \rho Q_{U} {g_1^{\prime}}^2 S^{\prime}\nonumber\\
\frac{d m_{\widetilde{b}_R}^2}{d t} &=& 4 \left(m_{\widetilde{t}_L}^2 + m_{H_d}^2 + m_{\widetilde{b}_R}^2 + \left|A_b\right|^2\right)
h_b^2\nonumber\\
                        &-& 8 \left( \frac{4}{3} g_3^2 \left|M_3\right|^2 +
                        \frac{1}{15} g_1^2 \left|M_1\right|^2 + \frac{1}{2} \rho Q_{D}^2 {g_1^{\prime}}^2
                        \left|M_1^{\prime}\right|^2\right)\nonumber\\
                        &+& \frac{2}{5} g_1^2 S + \rho Q_{D} {g_1^{\prime}}^2 S^{\prime}\nonumber\\
\frac{d m_{\widetilde{\tau}_L}^2}{d t} &=& 2 \left(m_{\widetilde{\tau}_L}^2 + m_{H_d}^2 + m_{\widetilde{\tau}_R}^2 +
\left|A_{\tau}\right|^2\right)
h_{\tau}^2\nonumber\\
                        &-& 8 \left( \frac{3}{4} g_2^2 \left|M_2\right|^2 + \frac{3}{20} g_1^2
                        \left|M_1\right|^2 + \frac{1}{2} \rho Q_{L}^2 {g_1^{\prime}}^2
                        \left|M_1^{\prime}\right|^2\right)\nonumber\\
                        &-& \frac{3}{5} g_1^2 S + \rho Q_{L} {g_1^{\prime}}^2 S^{\prime}\nonumber\\
\frac{d m_{\widetilde{\tau}_R}^2}{d t} &=& 4 \left(m_{\widetilde{\tau}_L}^2 + m_{H_d}^2 + m_{\widetilde{\tau}_R}^2 +
\left|A_{\tau}\right|^2\right)
h_{\tau}^2\nonumber\\
                        &-& 8 \left( \frac{3}{5} g_1^2
                        \left|M_1\right|^2 + \frac{1}{2} \rho Q_{E}^2 {g_1^{\prime}}^2
                        \left|M_1^{\prime}\right|^2\right)\nonumber\\
                        &+& \frac{6}{5} g_1^2 S + \rho Q_{E} {g_1^{\prime}}^2 S^{\prime}\nonumber\\
\frac{d m_{\widetilde{u}_L}^2}{d t} &=& - 8 \left( \frac{4}{3} g_3^2 \left|M_3\right|^2 + \frac{3}{4} g_2^2
                         \left|M_2\right|^2 +
                        \frac{1}{60} g_1^2 \left|M_1\right|^2 + \frac{1}{2} \rho Q_{Q}^2 {g_1^{\prime}}^2
                        \left|M_1^{\prime}\right|^2\right)\nonumber\\
                        &+& \frac{1}{5} g_1^2 S + \rho Q_{Q} {g_1^{\prime}}^2 S^{\prime}\nonumber\\
\frac{d m_{\widetilde{u}_R}^2}{d t} &=& - 8 \left( \frac{4}{3} g_3^2 \left|M_3\right|^2 +
                        \frac{4}{15} g_1^2 \left|M_1\right|^2 + \frac{1}{2} \rho Q_{U}^2 {g_1^{\prime}}^2
                        \left|M_1^{\prime}\right|^2\right)\nonumber\\
                        &-& \frac{4}{5} g_1^2 S + \rho Q_{U} {g_1^{\prime}}^2 S^{\prime}\nonumber\\
\frac{d m_{\widetilde{d}_R}^2}{d t} &=& - 8 \left( \frac{4}{3} g_3^2 \left|M_3\right|^2 +
                        \frac{1}{15} g_1^2 \left|M_1\right|^2 + \frac{1}{2} \rho Q_{D}^2 {g_1^{\prime}}^2
                        \left|M_1^{\prime}\right|^2\right)\nonumber\\
                        &+& \frac{2}{5} g_1^2 S + \rho Q_{D} {g_1^{\prime}}^2 S^{\prime}\nonumber\\
\frac{d m_{\widetilde{e}_L}^2}{d t} &=& - 8 \left( \frac{3}{4} g_2^2 \left|M_2\right|^2 + \frac{3}{20} g_1^2
                        \left|M_1\right|^2 + \frac{1}{2} \rho Q_{L}^2 {g_1^{\prime}}^2
                        \left|M_1^{\prime}\right|^2\right)\nonumber\\
                        &-& \frac{3}{5} g_1^2 S + \rho Q_{H_d} {g_1^{\prime}}^2 S^{\prime}\nonumber\\
\frac{d m_{\widetilde{e}_R}^2}{d t} &=& - 8 \left( \frac{3}{5} g_1^2
                        \left|M_1\right|^2 + \frac{1}{2} \rho Q_{E}^2 {g_1^{\prime}}^2
                        \left|M_1^{\prime}\right|^2\right)\nonumber\\
                        &+& \frac{6}{5} g_1^2 S + \rho Q_{E} {g_1^{\prime}}^2 S^{\prime}
\end{eqnarray}
where the corresponding RGEs in the MSSM are obtained by setting $g_1^{\prime} =0$, $h_s =0$ and $m_S^2 =0$
everywhere. Since U(1)$^{\prime}$ charges are family-universal the squared-masses of the right-handed sfermions
exhibit finite splitting only by their boundary values at $M_{GUT}$ (to the extent their Yukawa couplings can be
neglected). The beta functions of the scalar mass-squareds depend on $S$ defined in (\ref{seqn}), and on
\begin{eqnarray}
\label{speqn} S^{\prime} =\mbox{Tr}\left[m^2 Q\right] &=& 2 \left( Q_{H_u} m_{H_u}^2 + Q_{H_d} m_{H_d}^2 +
\frac{1}{2} Q_{S} m_{S}^2\right)\nonumber\\ &+& 6 Q_Q \left(m_{\widetilde{t}_L}^2 + m_{\widetilde{c}_L}^2 + m_{\widetilde{u}_L}^2\right) + 3
Q_U
\left(m_{\widetilde{t}_R}^2 + m_{\widetilde{c}_R}^2 + m_{\widetilde{u}_R}^2\right)\nonumber\\ &+& 3 Q_D \left(m_{\widetilde{b}_R}^2 +
m_{\widetilde{s}_R}^2 +
m_{\widetilde{d}_R}^2\right) + 2 Q_L \left(m_{\tau_L}^2 + m_{\mu_L}^2 + m_{e_L}^2\right)\nonumber\\ &+& Q_E
\left(m_{\widetilde{\tau}_R}^2 + m_{\widetilde{\mu}_R}^2 + m_{\widetilde{e}_R}^2\right)
\end{eqnarray}
which vanishes if mass-squareds are universal provided that $Z^{\prime}$-graviton-graviton anomaly cancels out
$i.e.$ $\mbox{Tr}\left[Q\right]= 0$.

\section*{Appendix B. Renormalization Group Equations in the NMSSM}
\setcounter{equation}{0}
\def\theequation{B.\arabic{equation}}
In this Appendix we list down RGEs for the NMSSM \cite{king}, for completeness. Since the model
exhibits no gauge extension with respect to MSSM, gauge couplings and gaugino
masses evolve precisely as in the MSSM (as mentioned below  equations
A.1 and A.4 in Appendix A). On the other hand, superpotential
parameters and soft masses are modified both in number and evolution,
and below we provide explicit expressions for their beta
functions.

The evolutions of the superpotential parameters are given by
\begin{eqnarray}
\frac{d h_t}{d t} &=& h_t \left( 6 h_t^2 + h_b^2 + h_s^2 - \frac{16}{3} g_3^2 - {3} g_2^2 - \frac{13}{15}
g_1^2\right)\nonumber\\
\frac{d h_b}{d t} &=& h_b \left( 6 h_b^2 + h_t^2 + h_{\tau}^2 + h_s^2 - \frac{16}{3} g_3^2 - {3} g_2^2 -
\frac{7}{15} g_1^2 \right)\nonumber\\
\frac{d h_{\tau}}{d t} &=& h_{\tau} \left( 4 h_{\tau}^2 + 3 h_b^2 + h_s^2 - {3} g_2^2 - \frac{9}{5} g_1^2\right)\nonumber\\
\frac{d h_s}{d t} &=& h_s \left( 4 h_s^2 + 3 h_t^2 + 3 h_b^2 + h_{\tau}^2 + \frac{1}{2} k_s^2 - 3 g_2^2 -
\frac{3}{5} g_1^2\right)\nonumber\\
\frac{d k_s}{d t} &=& k_s \left( 6 h_s^2 + \frac{3}{2} k_s^2\right)
\end{eqnarray}
so that $h_s$ and $k_s$ exhibit a correlated RG running yet rest of the Yukawas remain as in the U(1)$^{\prime}$
model (with $g_1^{\prime} = 0$, of course). The MSSM limit is achieved by putting $k_s =0$ and $h_s=0$ while
identifying $d \ln h_s /dt$ with $d \ln \mu /dt$.

The RG evolutions of the trilinear couplings are similar
\begin{eqnarray}
\frac{d A_t}{d t} &=& 2 \left( 6 h_t^2 A_t + h_b^2 A_b + h_s^2 A_s\right)\nonumber\\ &+&2 \left( \frac{16}{3}
g_3^2 M_3 + {3} g_2^2 M_2 + \frac{13}{15} g_1^2 M_1\right)\nonumber\\
\frac{d A_b}{d t} &=& 2 \left( 6 h_b^2 A_b + h_t^2 A_t + h_{\tau}^2 A_{\tau} + h_s^2 A_s\right)\nonumber\\ &+&2
\left( \frac{16}{3} g_3^2 M_3 + {3} g_2^2 M_2 + \frac{7}{15}
g_1^2 M_1\right)\nonumber\\
\frac{d A_{\tau}}{d t} &=& 2 \left( 4 h_{\tau}^2 A_{\tau} + 3 h_b^2 A_{b} + h_s^2 A_s \right)\nonumber\\ &+& 2
\left( {3} g_2^2 M_2 + \frac{9}{5}
g_1^2 M_1\right)\nonumber\\
\frac{d A_s}{d t} &=& 2 \left( 4 h_s^2 A_s + \frac{1}{2} k_s^2 A_k + 3 h_t^2 A_t + 3 h_b^2 A_b + h_{\tau}^2 A_{\tau} \right)\nonumber\\
&+& 2 \left( 3 g_2^2 M_2 + \frac{3}{5} g_1^2 M_1\right)\nonumber\\
\frac{d A_k}{d t} &=& 2 \left( 6 h_s^2 A_s + \frac{3}{2} k_s^2 A_k\right)
\end{eqnarray}
where the MSSM limit is obtained by putting $h_s=0$, $k_s =0$ everywhere and by identifying $A_s$ with the Higgs
bilinear mixing mass $B$.

Finally, the scalar mass-squared parameters run as follows
\begin{eqnarray}
\frac{d m_{H_u}^2}{d t} &=& 2 \left(m_{H_u}^2 + m_{H_d}^2 + m_{S}^2 + \left|A_s\right|^2\right) h_s^2\nonumber\\
                        &+& 6 \left(m_{H_u}^2 + m_{\widetilde{t}_L}^2 + m_{\widetilde{t}_R}^2 + \left|A_t\right|^2\right) h_t^2 \nonumber\\
                        &-& 8 \left( \frac{3}{4} g_2^2 \left|M_2\right|^2 + \frac{3}{20} g_1^2
                        \left|M_1\right|^2 \right)
                        + \frac{3}{5} g_1^2 S\nonumber\\
\frac{d m_{H_d}^2}{d t} &=& 2 \left(m_{H_u}^2 + m_{H_d}^2 + m_{S}^2 + \left|A_s\right|^2\right) h_s^2\nonumber\\
                        &+& 2 \left(m_{H_d}^2 + m_{\widetilde{\tau}_L}^2 + m_{\widetilde{\tau}_R}^2 + \left|A_{\tau}\right|^2\right)
h_{\tau}^2\nonumber\\
                        &+& 6 \left(m_{H_d}^2 + m_{\widetilde{t}_L}^2 + m_{\widetilde{b}_R}^2 + \left|A_b\right|^2\right) h_b^2 \nonumber\\
                        &-& 8 \left( \frac{3}{4} g_2^2 \left|M_2\right|^2 + \frac{3}{20} g_1^2
                        \left|M_1\right|^2\right)
                        - \frac{3}{5} g_1^2 S\nonumber\\
\frac{d m_{S}^2}{d t} &=& 4 \left(m_{H_u}^2 + m_{H_d}^2 + m_{S}^2 + \left|A_s\right|^2\right) h_s^2\nonumber\\
                         &+& \left(3 m_S^2 + \left|A_k\right|^2\right) k_s^2\nonumber\\
\frac{d m_{\widetilde{t}_L}^2}{d t} &=& 2 \left(m_{\widetilde{t}_L}^2 + m_{H_d}^2 + m_{\widetilde{b}_R}^2 + \left|A_b\right|^2\right)
h_b^2\nonumber\\
                        &+&2 \left(m_{\widetilde{t}_L}^2 + m_{H_u}^2 + m_{\widetilde{t}_R}^2 + \left|A_t\right|^2\right) h_t^2\nonumber\\
                        &-& 8 \left( \frac{4}{3} g_3^2 \left|M_3\right|^2 + \frac{3}{4} g_2^2 \left|M_2\right|^2\right)
                        + \frac{1}{5} g_1^2 S\nonumber\\
\frac{d m_{\widetilde{t}_R}^2}{d t} &=& 4 \left(m_{\widetilde{t}_L}^2 + m_{H_u}^2 + m_{\widetilde{t}_R}^2 + \left|A_t\right|^2\right)
h_t^2\nonumber\\
                        &-& 8 \left( \frac{4}{3} g_3^2 \left|M_3\right|^2 +
                        \frac{4}{15} g_1^2 \left|M_1\right|^2\right)
                        - \frac{4}{5} g_1^2 S\nonumber\\
\frac{d m_{\widetilde{b}_R}^2}{d t} &=& 4 \left(m_{\widetilde{t}_L}^2 + m_{H_d}^2 + m_{\widetilde{b}_R}^2 + \left|A_b\right|^2\right)
h_b^2\nonumber\\
                        &-& 8 \left( \frac{4}{3} g_3^2 \left|M_3\right|^2 +
                        \frac{1}{15} g_1^2 \left|M_1\right|^2\right)
                        + \frac{2}{5} g_1^2 S\nonumber\\
\frac{d m_{\widetilde{\tau}_L}^2}{d t} &=& 2 \left(m_{\widetilde{\tau}_L}^2 + m_{H_d}^2 + m_{\widetilde{\tau}_R}^2 +
\left|A_{\tau}\right|^2\right)
h_{\tau}^2\nonumber\\
                        &-& 8 \left( \frac{3}{4} g_2^2 \left|M_2\right|^2 + \frac{3}{20} g_1^2
                        \left|M_1\right|^2\right)
                        - \frac{3}{5} g_1^2 S \nonumber\\
\frac{d m_{\widetilde{\tau}_R}^2}{d t} &=& 4 \left(m_{\widetilde{\tau}_L}^2 + m_{H_d}^2 + m_{\widetilde{\tau}_R}^2 +
\left|A_{\tau}\right|^2\right)
h_{\tau}^2\nonumber\\
                        &-& 8 \left( \frac{3}{5} g_1^2
                        \left|M_1\right|^2\right)
                        + \frac{6}{5} g_1^2 S\nonumber\\
\frac{d m_{\widetilde{u}_L}^2}{d t} &=& - 8 \left( \frac{4}{3} g_3^2 \left|M_3\right|^2 + \frac{3}{4} g_2^2
                         \left|M_2\right|^2 +
                        \frac{1}{60} g_1^2 \left|M_1\right|^2\right)
                        + \frac{1}{5} g_1^2 S\nonumber\\
\frac{d m_{\widetilde{u}_R}^2}{d t} &=& - 8 \left( \frac{4}{3} g_3^2 \left|M_3\right|^2 +
                        \frac{4}{15} g_1^2 \left|M_1\right|^2\right)
                        - \frac{4}{5} g_1^2 S\nonumber\\
\frac{d m_{\widetilde{d}_R}^2}{d t} &=& - 8 \left( \frac{4}{3} g_3^2 \left|M_3\right|^2 +
                        \frac{1}{15} g_1^2 \left|M_1\right|^2\right)
                        + \frac{2}{5} g_1^2 S\nonumber\\
\frac{d m_{\widetilde{e}_L}^2}{d t} &=& - 8 \left( \frac{3}{4} g_2^2 \left|M_2\right|^2 + \frac{3}{20} g_1^2
                        \left|M_1\right|^2 \right)
                        - \frac{3}{5} g_1^2 S\nonumber\\
\frac{d m_{\widetilde{e}_R}^2}{d t} &=& - 8 \left( \frac{3}{5} g_1^2
                        \left|M_1\right|^2\right)
                        + \frac{6}{5} g_1^2 S
\end{eqnarray}
where evolution of $m_S^2$ is modified by the singlet cubic coupling whereas rest of the squared-masses run as in
the U(1)$^{\prime}$ model with $g_1^{\prime} = 0$.

\end{document}